\documentclass[11pt]{article}
\usepackage{xcolor}
\usepackage[english]{babel}
\usepackage[T1]{fontenc}
\usepackage[utf8]{inputenc}
\usepackage{authblk}
\usepackage{mathtools}
\usepackage{mathrsfs}
\usepackage{slashed}
\usepackage{amsmath,amssymb}
\usepackage{amsfonts}
\usepackage{graphicx,color}
\usepackage{subcaption}
\usepackage{comment}
\usepackage{wrapfig}
\usepackage{cite}
\usepackage{float}
\usepackage{hyperref}
\usepackage[capitalize]{cleveref}
\usepackage{setspace}
\usepackage[left=2.5cm,right=2.5cm,top=2.5cm,bottom=2.5cm]{geometry}

\hypersetup{
colorlinks=true,
linkcolor=blue,
filecolor=red,
urlcolor=blue,
}
\numberwithin{equation}{section}
\definecolor{refcol}{rgb}{0.95,0.1,0.1}
\hypersetup{colorlinks=false,linkcolor=blue,citecolor=red,urlcolor=blue,linktocpage}
\allowdisplaybreaks
\newcommand{\be}{\begin{equation}}
\newcommand{\ee}{\end{equation}}
\newcommand{\bea}{\begin{eqnarray}}
\newcommand{\eea}{\end{eqnarray}}

\def\XXint#1#2#3{{\setbox0=\hbox{$#1{#2#3}{\int}$ }
\vcenter{\hbox{$#2#3$ }}\kern-.6\wd0}}

\newcommand{\ket}[1]{\left| #1 \right\rangle}
\newcommand{\bra}[1]{\left\langle #1 \right|}
\newcommand{\cket}[1]{\left| #1 \right)}
\newcommand{\cbra}[1]{\left( #1 \right|}
\newcommand{\call}{\mathcal{L}}
\newcommand{\calo}{\mathcal{O}}
\newcommand{\cala}{\mathcal{A}}
\newcommand{\norm}[1]{||#1 ||}
\newcommand{\calh}{\mathcal{H}}

\newcommand{\cbraket}[2]{\left( #1 \middle| #2 \right)} 

\makeatletter
\let\old@maketitle\@maketitle
\def\@maketitle{%
  \begingroup
    \let\newpage\relax 
    \begin{flushright}
    \end{flushright}
    \old@maketitle
  \endgroup
}
\makeatother

\begin{document}

	\begin{titlepage}
		\thispagestyle{empty}
		\title{ 
			{\huge\bf Symmetry-resolved Krylov Complexity and the \\
Uncoloured Tensor Model}
		}
		\bigskip
		\author{
			{\bf Shaliya Kotta${}^{}$}\thanks{{\tt shaliya\_p240183ph@nitc.ac.in,shaliyakotta@gmail.com}}
			{\bf ~and P N Bala Subramanian${}^{}$}\thanks{{\tt pnbala@nitc.ac.in}}\\
			\smallskip\hfill\\      
			{\small
				{\it Department of Physics, National Institute of Technology Calicut,}\\
{\it Kozhikode, Kerala - 673601, India}				
			}
		}
		
		\bigskip\bigskip\bigskip\bigskip
		\vfill
		\date{		
			\begin{quote}
				\centerline{{\bf Abstract}}
				{\small
					The symmetry-resolved Krylov complexity is a useful tool for studying the chaotic properties of systems endowed with symmetries. We investigate the conditions under which an invariant Hermitian operator would have the symmetry-resolved Krylov complexity in a charge subspace identical to the Krylov complexity of the full operator, and find a necessary and sufficient condition that ensures this matching at all temperatures. Further, we study the Krylov complexity of the Uncoloured Tensor Model, a disorder-free kin of the SYK Model, which has a plethora of symmetries. We identify charge subspaces of the same operator in which equipartition holds, as well as those where it doesn't, along with observations on the underlying mechanism behind these. We also find that, within the computational limits, the Krylov complexity, averaged over the symmetry subspace, is bounded above by that of the operator in the full space. }
			\end{quote}
            }


	
\end{titlepage}
\thispagestyle{empty}
\maketitle\vfill \eject

\tableofcontents
\bigskip
\hrule


\section{Introduction}
It is fascinating that despite a century of work studying Quantum Physics, fundamental questions about its nature remain as relevant as they were at its inception. One of the key questions that has remained at the center of attention has been the emergence of classicality, in different guises -- implications of quantum physics in black hole dynamics: microscopic nature of the Bekenstein-Hawking entropy\cite{Bekenstein:1973ur, Hawking:1982dh,tHooft:1993dmi, Susskind:1994vu} and the Information Paradox\cite{Hawking:1975vcx}, ergodic nature of quantum systems and the emergence of classical statistical mechanics\cite{neumann1929beweis,2006NatPh...2..754P, Srednicki_1994}, quantum decoherence \cite{Zeh1970, Schlosshauer2007decoherence}, to name a few. It can also be seen that these questions span disciplines and have been investigated in very similar systems from different perspectives. The advent of the AdS/CFT correspondence \cite{Maldacena:1997re, Witten:1998qj} has played a crucial role in this by revealing connections of gravitational systems to those studied in condensed matter literature \cite{Herzog_2007}, most recently that of black holes and non-Fermi liquids via the SYK Model \cite{PhysRevLett.70.3339_SYKintro, Kitaev_lectureonline} and Tensor Models \cite{2016arXiv161009758W_edwardwitten, Gurau_2017, PhysRevD.95.046004_uncolored_model_into}.

Quantum chaos started gaining traction in the study of black hole physics with the SYK Model saturating the bounds of growth of the Out-of-Time Ordered Correlation functions (OTOCs) set by the study of shocks in black hole backgrounds \cite{Shenker:2013pqa, Maldacena2015waaAboundonchaos}. Further studies revealed the connection of these quantum models to the Jackiw-Teitelboim gravity in the low-energy limit\cite{maldacena2016conformalsymmetrybreakingdimensional}, as well as the emergence of conformal behaviour. While the SYK model was the central player in these studies, other \emph{fermionic} quantum systems emerged, which were devoid of disorder averaging, starting with the {\it Coloured} Tensor Models of Gurau-Witten \cite{2016arXiv161009758W_edwardwitten, Gurau_2017} and the {\it Uncoloured} Tensor Models of Klebanov-Tarnopolsky \cite{PhysRevD.95.046004_uncolored_model_into}, which exhibited the same melonic diagram dominance in the large-N limit (melonic dominance of $O(N)$ random tensors was first shown in \cite{Carrozza:2015adg}). Even in the finite-N limit, they have been shown to have interesting properties such as indications of chaos as seen from spectral probes such as Spectral Form Factor and Level-Spacing Distributions\cite{Krishnan2017bala_colored, Krishnan2017_uncolored, Krishnan2018_conrastingSYKlike, Itoyama_2017, Choudhury_2018, Chaudhuri_2018, Carrozza_2018}, exact solvability\cite{Krishnan_2017, PhysRevLett.120.201603, Klebanov_2018} and connections to combinatorics\cite{Geloun_2017, pavan}. 

Following the Operator Growth Hypothesis\cite{PhysRevX.9.041017_universalopgrowthhypothesis_parkerEtAl}, the study of chaos has been focused on the study of Krylov complexity of the operators, whereby the growth of the Lanczos coefficients in the Heisenberg evolution of the operator is used to decipher information about the Lyapunov exponent of the system. In particular, the quantum bound on the Lyapunov exponent manifests itself as a bound on the rate of growth of the Lanczos coefficients. Systems that are maximally chaotic saturate these bounds, as was shown in the case of the SYK Model, Virasoro model \cite{Caputa:2021sib}, among others \cite{Hashimoto_2023, Baggioli_2025, Avdoshkin_2024, PhysRevD.109.046017, Caputa_2024} (see \cite{Nandy_2025,rabinovici2025krylovcomplexity} for comprehensive reviews and the references therein). Further, models such as SYK and their deformations\cite{Bhattacharjee_2023,chryssanthacopoulos2026krylovcomplexitysupersymmetricsyk}, spin chains with integrability-breaking interactions\cite{Yates_2020,Noh_2021,Rabinovici_2022,Bhattacharyya_2023,Bhattacharya_2024}, etc. have all been subject to extensive numerical study. While it is possible to study the Krylov complexity of certain systems analytically, at least in certain regimes, numerical computations are often used, especially for spin systems on lattices. Apart from constraints arising from precision issues in the orthonormalisation of the vectors, the exponential scaling of the matrix size in fermionic systems with increasing $N$ severely limits the number of degrees of freedom that can be included in the study. In this context, the symmetry-resolved Krylov complexity\cite{9v9v-54zv_blockdiagKrylov,Caputa_2026} provides a path for simplifying the calculation of Krylov complexity, if we can identify a symmetry subspace in which the symmetry-resolved Krylov complexity of the projected operator is identical to the Krylov complexity of the full operator under consideration, which we call \textit{equipartition}. Such a reduction in computational cost could be utilised to study larger systems while preserving the information we seek. However, an operator will have to satisfy additional conditions, in addition to possessing the symmetry under consideration, for equipartition to hold. We systematically derive these conditions and examine constructive ways to employ them. Further, we do an explicit numerical study on the Klebanov-Tanopolsky (uncoloured) Tensor Model \cite{PhysRevD.95.046004_uncolored_model_into}, closely related to the Gurau-Witten (coloured) Tensor Model, which exhibits a very large level of degeneracy at each energy level\cite{Krishnan2018_conrastingSYKlike}, where we analyse different symmetries that do and don't admit equipartition and test the conditions that we find. Finally, we also comment on peculiar issues encountered in the use of the Lanczos algorithm\cite{Paige1971_thesis_errorLanczos, Parlett1979_Lanczos_SO, Simon1984_Lanczos_PRO,k94p-vls8_error_Escaping_Krylovspace}, due to the large level of degeneracy and the significant size of the operator matrices.

The organisation of the paper is as follows. In Section \ref{sec: setup}, we review the key aspects of Krylov complexity and its symmetry-resolved counterpart. In Section \ref{sec: conditions}, we study the conditions under which we can expect the Krylov complexity to be the same in the full space and a charge subspace, and we put it to the test in the context of the Klebanov-Tarnopolsky Tensor Model in Section \ref{sec: tensor model}. We conclude and give general remarks in Section \ref{sec: conclusions}. Additional computational details for Section \ref{sec: conditions} are provided in Appendix \ref{appndx: proof equipartition}, numerical precision issues are discussed in Appendix \ref{appndx: numerical_instability}, and the symmetries of the Tensor Model are reviewed in Appendix \ref{appndx: symmetries of H}.

\section{Krylov Complexity}\label{sec: setup}
Krylov complexity was first introduced in \cite{PhysRevX.9.041017_universalopgrowthhypothesis_parkerEtAl} as a tool to study the chaotic behaviour of quantum systems, as a bound on a class of 'q-complexities', that include notions such as OTOCs and operator complexity. Krylov complexity quantifies the growth of an initial simple operator ($\mathcal{O}$) in an operator basis, called the \textit{Krylov basis}, as it evolves under the Heisenberg equations of motion. For $\mathcal{O}$, this amounts to exploring the operator space, under time evolution, by increasing contributions coming from larger nested commutators $[H,[H,[\dots[H,\calo]\dots]]] $. We briefly review the relevant details below.

Consider a quantum system with Hamiltonian $H$, and a Hermitian operator $\mathcal{O}$. In the Heisenberg picture, the time evolution of $\mathcal{O}$ is given by,
\begin{equation}
    \mathcal{O}(t) = e^{-iHt} \mathcal{O}(0) e^{iHt} = e^{i\mathcal{L}t} \mathcal{O}(0),
\end{equation}
where $\mathcal{L}=[H, \cdot \,]$ is the Liouvillian superoperator. The set of all (linear) operators in the Hilbert space itself forms a vector space $\calo(\calh)$. Let's denote the operator $\mathcal{O}$, an element of this operator-vector space, by $\cket{\mathcal{O}}$. The time evolution of the operator is then given by
\begin{equation}
    \cket{\mathcal{O}(t)} = e^{i\mathcal{L}t} \cket{\mathcal{O}} = \sum_{n=0}^{\infty}\frac{(i\,t)^n}{n!}\mathcal{L}^n \, \cket{\mathcal{O}}.
\end{equation}
With the choice of the finite temperature Wightman inner product
\begin{equation}
    \cbraket{\mathcal{A}}{\mathcal{B}}_\beta= \frac{1}{Z}\text{Tr}(e^{-\beta H/2}\,\mathcal{A}^\dagger\,e^{-\beta H/2}\,\mathcal{B}),
\end{equation}
where $Z=\text{Tr}(e^{-\beta H})$, and the Lanczos algorithm (see Appendix \ref{appndx: numerical_instability}), we obtain the Lanczos coefficients $b_n$, and the orthonormalised basis in the Krylov space $\{\cket{K_n} \}$.

One can think of the Krylov basis as forming a chain of vectors, numbered 0 to $K-1$. Expanding $\cket{\mathcal{O}(t)}$ in the Krylov basis, we get
\begin{align}
    \cket{\mathcal{O}(t)} &= \sum_{n}^{K-1} i^n \phi_n(t) \,\cket{K_n}, \ \  \text{where  } \phi_n(t)=i^{-n} \cbraket{K_n}{\mathcal{O}(t)}_\beta.
\end{align}
The evolution of $\mathcal{O}(t)$ then becomes a discrete Schr\"odinger equation for evolution of wave function $\phi_n(t)$
\begin{equation}
    \partial_t\phi_n(t) = b_n\phi_{n-1}-b_{n+1}\phi_{n+1}.
\label{eq:phi_n Schroedinger eq}
\end{equation}
Krylov Complexity is defined as the average position on the chain
\begin{equation}
    C_K(t)= \sum_{n=0}^{K-1} n\,|\phi_n(t)|^2,
\label{eq:Ck in terms of phi_n}
\end{equation}
which is also the expectation value of the superoperator
\begin{equation}
    \widehat{\mathcal{C}_K}=\sum_{n=0}^{K-1} n \cket{K_n}\cbra{K_n},
\end{equation}
where each Krylov vector $\cket{K_n}$ is an eigenvector of $\widehat{\mathcal{C}_K}$ with eigenvalue $n$.

\subsection{Symmetry-resolved Krylov Complexity}\label{subsec: sym-resolved-kcomp}
\textit{Symmetry-resolved} Krylov complexity was introduced in \cite{9v9v-54zv_blockdiagKrylov} for invariant operators in systems that are endowed with symmetries. It was shown that the average symmetry-resolved Krylov complexity exactly captures the Krylov complexity of the full system at early times. However, the late-time behaviour is a complex interplay between different charge sectors, due to which the Krylov complexity of the full operator fails to be a simple function of that of different charge sectors, and it was conjectured that the averaged symmetry-resolved Krylov complexity is bounded above by the Krylov complexity of the full operator.

Let $Q$ be a conserved charge ($[H,Q]=0$), where $Q$ generates a $U(1)$ symmetry. Then the Hilbert space has a natural decomposition as a direct sum of the charge sectors of $Q$: $\mathcal{H}=\oplus_q \mathcal{H}_q$, where $q$ is the eigenvalue of $Q$. Consider an operator $\calo$ that commutes with $Q$ (invariant operator) but not with $H$, then in the eigenbasis of $Q$, both $H$ and $\calo$ are block diagonal,
\begin{align}
    &\calo(t)=\sum_q \calo_q(t) =\sum_q \Pi_q\, \calo(t), \ \text{and    } H =\sum_q H_q=\sum_q \Pi_q H,
\end{align}
where $\Pi_q$ is the projector onto subspace $\mathcal{H}_q$. Applying the Lanczos algorithm on the block operators individually, we get the Krylov basis $\left\{\cket{K_n^{(q)}}\,| n=0,1,...\mathcal{K}_q-1\right\}$ of dimension $\mathcal{K}_q$. Assuming that the operators $\cket{\calo_{q}(0)}$ of all blocks and the full operator $\cket{\calo(0)}$ are normalised, we get
\begin{align}
    \cket{\calo(t)} &= \sum_q \sqrt{p_q} \cket{\calo_q(t)},\\
    \text{where} \ \ p_q&= \frac{1}{Z}\text{Tr}(e^{-\beta H/2}\,\calo_q^\dagger\,e^{-\beta H/2}\,\calo_q),
\end{align}
with $\sum_q p_q =1$. This allows $p_q$ to be interpreted as the probability associated with each charge sector for the operator. With $\phi_n^{(q)}(t)$ being the component of $\cket{\calo_q(t)}$ along $\cket{K_n^{(q)}}$, symmetry-resolved Krylov complexity is defined as
\begin{equation}
    C_K^{(q)}(t)=\sum_0^{\mathcal{K}_q-1} n |\phi_n^{(q)}(t)|^2,
\end{equation}
and the average Krylov complexity over the sectors is defined as
\begin{equation}
    \overline{C}(t)=\sum_q p_q C_K^{(q)} (t).
\end{equation}
In \cite{9v9v-54zv_blockdiagKrylov}, it was shown that for small times
\begin{equation}
    C_K(t)-\overline{C}(t)= \left( \, \sum_q p_q \left(b_1^{(q)} \right)^4-b_1^4\right)\frac{t^4}{2} + O(t^6).
\end{equation}
Thus, the dynamics of the full operator are captured by averaging the dynamics of each block as if the other blocks were absent. Further, it was conjectured in \cite{9v9v-54zv_blockdiagKrylov} that $C_k(t)-\overline{C}(t) \geq 0$ for all $t$, motivated by the mixing of sectors in Krylov vectors, since for $n\geq 2$, $\cket{K_n}$ cannot be written solely in terms of the $n^{\text{th}}$ Krylov vector of the charge sectors.

\section{Conditions for equipartition } \label{sec: conditions}

The symmetry-resolved Krylov complexity can be leveraged to reduce the computational load if we can pin down the structures responsible for \textit{equipartition}. The conjecture of \cite{9v9v-54zv_blockdiagKrylov} is that the average Krylov complexity is bounded above by that of the operator in the full Hilbert space, for a general operator that respects the symmetry of the Hamiltonian under consideration. It is also possible to have equipartition of complexity, where the growth of the operators in each charge subspace matches that in the full space. In this case, it would be possible to find the smallest sector where such a condition is met, and use that to compute the complexity of the operator in the full space. Here, we would like to obtain the most general conditions under which such equipartition holds.

Let us start with a basis for the Hilbert space labelled by the eigenvalues of the commuting operators $ H $ and $ Q $, \[ \Big\{ \ket{E_a,q,\rho} \Big|~a=1,2,\dots, N; ~ q=1,2,\dots,\dim{Q_a}; ~ \rho\in \mathcal{A}_{(a,q)} \Big\}, \] where $\mathcal{A}_{(a,q)} $ is some indexing set that depends on both $(a,q)$ in general. The basis of operators is then $ \Big\{ \ket{E_a,q,\rho}\bra{E_b,q',\sigma} \Big\}$, which are also eigenstates of the Liouvillian with eigenvalue $(E_a-E_b) = \omega_{ab}$. An operator that respects the symmetry under consideration is restricted to the $q=q'$ subspace.

At the outset, it is not a given that the range of $q$ values, denoted by $|Q_a| $, is the same for all the $a$ subspaces, nor that $ |\cala_{(a,q)}| $ is the same in all the $(a,q) $ subspaces. We take $|Q| = \text{max}(|Q_a|)$, and if an $a$ is void of a particular $q$ value, then $\cala_{(a,q)} = \varnothing $. An operator that respects the symmetry under consideration could be expanded as
\begin{align}\label{eq:opexpand}
    \calo = \sum\limits_{a=1}^{N}\sum\limits_{b=1}^{N}\sum\limits_{q=1}^{|Q|}\sum\limits_{\rho = 1}^{|\cala_{(a,q)}|}\sum\limits_{\sigma = 1}^{|\cala_{(b,q)}|} c_{a,b,q}^{\rho,\sigma} \, \ket{E_a,q,\rho}\bra{E_b,q,\sigma}.
\end{align}
We will ignore the trivial case where $\calo$ has a non-trivial projection over just a single $q$-subspace in what follows, as the Krylov complexity in the full space and that particular $q$-subspace will be the same by construction, and focus on operators that have projections onto two or more $q$-subspaces. We start by picking a particular subspace, say $q_0$, and demanding that the Lanczos coefficients and thereby the Krylov complexity are the same for $\calo $ and $\calo_{q_0}$. For this to be true, $b_n = b_n^{(q_0)} $ for all values of $n$, which requires the Krylov space vectors to satisfy the condition 
\begin{align}
    \dfrac{\norm{\call^n\calo}_\beta^2}{\norm{\calo}_\beta^2} = \dfrac{\norm{\call^n\calo_{q_0}}_\beta^2}{\norm{\calo_{q_0}}_\beta^2}.
\end{align}
If we demand that this condition is met for all values of $\beta$, the operator is then required to satisfy (see Appendix \ref{appndx: proof equipartition} for details)
\begin{align}\label{eq:qmatching}
    \dfrac{\sum\limits_{\rho = 1}^{|\cala_{(a,q_0)}|} \sum\limits_{\sigma = 1}^{|\cala_{(b,q_0)}|} |c_{a,b,q_0}^{\rho,\sigma}|^2} {\sum\limits_{q'=1}^{|Q|}\sum\limits_{\rho = 1}^{|\cala_{(a,q')}|} \sum\limits_{\sigma = 1}^{|\cala_{(b,q')}|} |c_{a,b,q'}^{\rho,\sigma}|^2} = \eta_{q_0} ,
\end{align}
for all $a,b$, where $\eta_{q_0}$ is some constant. This means that the ratio of the norm of the operator in the $(a,b,q_0)$-subspace to that in the $(a,b)$-subspace is independent of $a,b$. Suppose we consider the complementary subspace to $q_0$, labelled by $\overline{q_0}=\{q|q\in Q\backslash q_0 \}$, we have 
\begin{align}
    \dfrac{\sum\limits_{q'\in\, \overline{q_0}}\sum\limits_{\rho = 1}^{|\cala_{(a,q_0)}|} \sum\limits_{\sigma = 1}^{|\cala_{(b,q_0)}|} |c_{a,b,q'}^{\rho,\sigma}|^2} {\sum\limits_{q'=1}^{|Q|}\sum\limits_{\rho = 1}^{|\cala_{(a,q')}|} \sum\limits_{\sigma = 1}^{|\cala_{(b,q')}|} |c_{a,b,q'}^{\rho,\sigma}|^2} = 1- \eta_{q_0} ,
\end{align}
which results in the Lanczos coefficients of $\calo_{(\overline{q}_0)}$ being the same as those of $\calo $ ($b_n = b_n^{(\overline{q_0})}$), and subsequently leads to identical Krylov complexity. It is possible to apply this iteratively within the $\overline{q}_0$ subspace to simplify it further.

We could rearrange the terms in (\ref{eq:qmatching}) to fix the ratios of norms in an $(a,b)$-subspace, to obtain 
\begin{align}\label{eq:abmatching}
    \dfrac{\sum\limits_{\rho = 1}^{|\cala_{(a,q_0)}|} \sum\limits_{\sigma = 1}^{|\cala_{(b,q_0)}|} |c_{a,b,q_0}^{\rho,\sigma}|^2} {\sum\limits_{a,b=1}^{N}\sum\limits_{\rho = 1}^{|\cala_{(a,q_0)}|} \sum\limits_{\sigma = 1}^{|\cala_{(b,q_0)}|} |c_{a,b,q_0}^{\rho,\sigma}|^2} = \dfrac{\sum\limits_{q'=1}^{|Q|}\sum\limits_{\rho = 1}^{|\cala_{(a,q')}|} \sum\limits_{\sigma = 1}^{|\cala_{(b,q')}|} |c_{a,b,q'}^{\rho,\sigma}|^2}{\sum\limits_{a,b=1}^{N}\,\sum\limits_{q'=1}^{|Q|}\sum\limits_{\rho = 1}^{|\cala_{(a,q')}|} \sum\limits_{\sigma = 1}^{|\cala_{(b,q')}|} |c_{a,b,q'}^{\rho,\sigma}|^2}= \zeta_{a,b}, 
\end{align}
where $\zeta_{a,b}$ depends solely on $a$ and $b$, since the term in the middle is manifestly $q_0$-independent. Thus, we finally arrive at the condition 
\begin{align}\label{eq: coeffratio}
    \dfrac{\sum\limits_{\rho = 1}^{|\cala_{(a,q_0)}|} \sum\limits_{\sigma = 1}^{|\cala_{(b,q_0)}|} |c_{a,b,q_0}^{\rho,\sigma}|^2}{\sum\limits_{a,b=1}^{N}\,\sum\limits_{q'=1}^{|Q|}\sum\limits_{\rho = 1}^{|\cala_{(a,q')}|} \sum\limits_{\sigma = 1}^{|\cala_{(b,q')}|} |c_{a,b,q'}^{\rho,\sigma}|^2} &= \eta_{q_0}\, \zeta_{a,b}\,.
\end{align}

Recasting these statements concisely and basis-independent, (\ref{eq:abmatching}) and (\ref{eq: coeffratio}) take the form
\begin{align}\label{eq: gen_equipartition}
    \dfrac{\text{Tr}\left(\Pi_{b,q_0} \calo^\dagger \Pi_{a,q_0} \calo  \right)}{\text{Tr}\left(\Pi_{q_0} \calo^\dagger \calo  \right)} &=     \dfrac{\text{Tr}\left(\Pi_{b} \calo^\dagger \Pi_{a} \calo  \right)}{\text{Tr}\left(\calo^\dagger \calo  \right)} \\
   \text{and} \ \  \text{Tr}\left(\Pi_{b,q_0} \calo^\dagger \Pi_{a,q_0} \calo  \right) &= \eta_{q_0} \zeta_{a,b} \text{Tr}(\calo^\dagger \calo)
\end{align}

An obvious condition for the non-existence of equipartition, as seen from above, is when the operator has $\Pi_{a,q_0}\calo\Pi_{b,q_0} = 0 $ while $\Pi_{a}\calo\Pi_{b} \neq 0 $ for at least one pair $(a,b)$. If it is known, from other arguments or numerically, that the operator meets this criterion, the symmetry-resolved Krylov complexity in the $q_0$ subspace will be distinct from that in the full space.

\subsection{A Constructive Approach for Identifying Symmetry-Resolved Sectors}\label{subsec: constructive_method}

Directly verifying the condition in (\ref{eq: gen_equipartition}) for equipartition requires knowledge of the full eigensystem of the Hamiltonian. This does not significantly simplify the computation, as the eigensystem must be computed numerically. However, we can leverage the system's symmetry algebra to derive much simpler conditions that ensure equipartition, which we discuss below. It is to be noted that the conditions we obtain below are merely sufficient and are not necessary. However, in systems that satisfy the conditions listed below, equipartition is assured, and no additional verification is required. 

Let us start with a conserved charge $Q$, with only two eigenvalues $\pm q$. The equipartition of Krylov complexity among the blocks $\calo_{+q}$ and $\calo_{-q}$ of an invariant operator is guaranteed if there exists a unitary operator $A$ that satisfies the following conditions:
\begin{align}
    [A, H] &= 0, \label{eq:cond_A1} \\
    \{A, Q\} &= 0, \label{eq:cond_A2} \\
    [A, \calo]_{\pm} &= 0. \label{eq:cond_A3}
\end{align}
Here, equation (\ref{eq:cond_A2}) ensures that $A$ maps the $q$ sector to the $-q$ sector via $\Pi_{-q} = A \Pi_{q} A^\dagger $, and we get
\begin{align}
    \text{Tr}(\Pi_{b,-q}\calo^\dagger\Pi_{a,-q}\calo) = \text{Tr}(A \Pi_{b,q} A^\dagger\calo^\dagger A \Pi_{a,q} A^\dagger\calo) = \text{Tr}(\Pi_{b,q}\calo^\dagger\Pi_{a,q}\calo)
\end{align}
where equation (\ref{eq:cond_A3}) and cyclicity of the trace are used to move around $A$. This ensures that the condition is met for both $\pm q$ and, consequently, equipartition.

This construction naturally generalises to a further decomposition via a second charge. Suppose that there exists a second conserved charge $P$ that commutes with $Q$, with eigenvalues $\pm p$. Then, $\calo_q$ decomposes into $\calo_{q,p}$ blocks, with a further equipartition among the $\pm p$ blocks if there exists a second unitary operator $B$ such that the following conditions are satisfied:
\begin{align}
    [P, Q] &= 0, & [P, A] &= 0,\label{eq:cond_B1} \\
    [B, H] &= 0, & [B, Q] &= 0,\label{eq:cond_B2} \\
    \{B, P\} &= 0, & [B, \calo]_{\pm} &= 0.\label{eq:cond_B3}
\end{align}
Thus, the structure guarantees equipartition of Krylov complexity across the four blocks of the total charge $Q_{\text{tot}} = Q + P$, with eigenvalues $\pm q \pm p$.

In the case where the symmetry charge has more than two eigenvalues, it becomes harder to obtain conditions on the algebra of the operators that can be meaningfully used. From the discussion above, it is clear that for any two charge sectors $q_i$ and $q_j$, if we can find a unitary operator that commutes with $H$ and satisfies 
\begin{align}\label{eq: ij_matching}
\Pi_{q_i} = A \Pi_{q_j} A^\dagger, \ \text{and } \ [A,\calo]_\pm =0,
\end{align}
we can show that the two sectors have the same Lanczos coefficients, because 
\begin{align}
    \dfrac{\text{Tr}\left(\Pi_{b,q_i} \calo^\dagger \Pi_{a,q_i} \calo  \right)}{\text{Tr}\left(\Pi_{q_i} \calo^\dagger \calo  \right)} = \dfrac{\text{Tr}\left(\Pi_{b,q_j} \calo^\dagger \Pi_{a,q_j} \calo  \right)}{\text{Tr}\left(\Pi_{q_j} \calo^\dagger \calo  \right)} . 
\end{align}

\subsection{Comments on Noether Symmetries}
If the Hamiltonian possesses a Noether symmetry, the Hilbert space of the system decomposes into irreducible representations of the algebra. For concreteness, let us assume the continuous symmetry to be SU(2), although the discussion can be generalised to any Lie algebra. The Hilbert space, in this case, has the structure 
\begin{align}
    \calh  = \bigoplus_j \mathcal{W}_j \otimes V_j,
\end{align}
where $V_j$ are the irreducible representations labelled by $j$, and $\mathcal{W}_j$ are the multiplicity spaces associated with each $j$ which in general could all be distinct. A convenient basis for the Hilbert space is
\begin{align}\label{eq: basisonh}
    \mathcal{B}(\calh) = \{ \ket{E_a,\rho,j}\otimes \ket{j,m} \}.
\end{align}

The space of operators will then be understood in terms of these spaces as 
\begin{align}
    \calo(\calh)= \bigoplus_{j,j'} (\mathcal{W}_j\otimes \mathcal{W}_{j'}^*) \otimes (V_j\otimes V_{j'}^*) = \bigoplus_{k} \Theta_k \otimes \mathcal{T}^k,  
\end{align}
where the $\mathcal{T}^k$ are spherical tensor operators of rank $k\in \{0,1,\dots, 2j_{max}\}$. Any operator $\calo$ can be expanded as a linear combination of spherical tensors,
\begin{align}
    \calo &= \sum_{k,q} \theta_{k,q} \otimes \tau^k_q ,
\end{align}
and expressing it in the basis obtained from (\ref{eq: basisonh}), we get
\begin{align}
    \calo &= \sum\limits_{k,q}\sum\limits_{j,j',m,m'}\sum\limits_{a,b,\rho,\sigma} F_{a,b,\rho,\sigma}^{j,j',k,q} \ket{E_a,\rho,j}\bra{E_b,\sigma,j'}\otimes G_{j,j',k,q}^{m,m'}\ket{j,m}\bra{j',m'},\\
     \text{where    } F_{a,b,\rho,\sigma}^{j,j',k,q} &= \bra{E_a,\rho,j}\theta_{k,q}\ket{E_b,\sigma,j'}, \ \ G_{j,j',k,q}^{m,m'} = \bra{j,m}\tau^k_q\ket{j',m'}.
\end{align}
This is nothing but the Wigner-Eckart theorem, written in a manner that explicitly tracks the multiplicity space associated with the irreducible representations.

We restrict further discussions to spherical tensors that commute with $J_z$, whereby only $m=m'$ terms survive; we are looking for conditions for equipartition, and an operator that mixes up the $J_z$ eigenvalues is of no utility in this context. For equipartition, we require a factorisation of the subspace norms 
\begin{align}
    l^2_{(a,b),m} = \sum\limits_{k,k'}\sum\limits_{j,j'}\sum\limits_{\rho,\sigma} \Big(F_{a,b,\rho,\sigma}^{j,j',k} G_{j,j',k,0}^m\Big)^*  \, \Big(F_{a,b,\rho,\sigma}^{j,j',k'} G_{j,j',k',0}^m\Big)= \eta_m \,\zeta_{(a,b)}\, \norm{\calo}^2_{\beta=0},
\end{align}
which, however, does not explicitly impose a complete restriction on the factorisation of the coefficients themselves in terms of $a$, $b$ and $m$. There could be operators where the factorisation results from cancellations across different terms. 

We could, though, work out sufficient conditions that ensure factorisation at the cost of greater restrictiveness, which could aid in the bottom-up construction of the operators. The first assumption is $ F_{a,b,\rho,\sigma}^{j,j',k} = F_{a,b,\rho,\sigma}^{j,j'}$, i.e. independent of $k$, whereby we get
\begin{align}
     l^2_{(a,b),m} = \sum\limits_{j,j'} \Big(\sum\limits_{\rho,\sigma}( F_{a,b,\rho,\sigma}^{j,j'}){}^* \,\, F_{a,b,\rho,\sigma}^{j,j'}\Big) \Big(\sum\limits_{k,k'}  (G_{j,j',k,0}^m){}^*\,\,  G_{j,j',k',0}^m\Big).
\end{align}
To get around the mixing of terms under the sum in $j,j'$, we could look at two possibilities: (i) the only non-trivial coefficient is for $F_{a,b,\rho,\sigma}^{j,j'}\propto \delta^j_{j_0}\delta^{j'}_{j_0'}$, (ii) $F_{a,b,\rho,\sigma}^{j,j'}$ are independent of $j,j'$. A particularly useful case that falls into the first category is when the Hilbert space is composed of a single irreducible representation, and the operator is of the form $\theta \otimes (\sum_k \tau^k_0)$.
Let us analyse the second condition in a bit more detail, as the conditions for equipartition are dependent on the structure of the multiplicity space of the irreducible representations in the Hilbert space. If the Hilbert space is such that the multiplicity space is identical across all irreducible representations, by which we mean the spectrum of the Hamiltonian in all the multiplicity spaces is the same, $\calh =  \mathcal{W}\otimes (\oplus_j V_j )$, whereby $\calo(\calh) = (\mathcal{W}\otimes \mathcal{W}^*)\otimes \oplus_{j,j'} (V_j\otimes V_{j'}^*)$, and the operator is constructed by taking a tensor product of the form $\theta \otimes (\sum_k \tau^k_0)$.  In the case where the multiplicities are not all identical across different representations, one condition where we get the required factorisation is when the operators fall in the subspace 
\begin{align}
    (\widehat{\mathcal{W}}\otimes\widehat{\mathcal{W}}^*) \otimes \left(\bigoplus_{j,j'} (V_j\otimes V_{j'}^*) \right) \subset \bigoplus_{j,j'} (\mathcal{W}_j\otimes \mathcal{W}_{j'}^*) \otimes (V_j\otimes V_{j'}^*)
\end{align}
for some $\widehat{\mathcal{W}}\subset \mathcal{W}_j\, \forall j$. This, however, is generally not expected for a dense operator and would require fine-tuned cancellations to achieve equipartition. 

\section{Complexity in the Uncoloured Tensor Model}\label{sec: tensor model}
Fermionic Tensor Models were first introduced by Edward Witten in \cite{2016arXiv161009758W_edwardwitten} as a class of 0+1 dimensional quantum mechanical models that have a large $N$ limit similar to the famous SYK Model \cite{PhysRevLett.70.3339_SYKintro, Kitaev_lectureonline}, but do not require the disorder. Later, a class of \textit{Uncoloured} Tensor Models with a smaller symmetry group than their Coloured counterpart were introduced in \cite{PhysRevD.95.046004_uncolored_model_into}. The SYK Model shows chaotic behaviour in studies of OTOCs\cite{PhysRevX.9.041017_universalopgrowthhypothesis_parkerEtAl}, spectral statistics \cite{Maldacena_2016_commentsonthesykmodel_ig:OTOC, Cotler2017_Blackholesandrandommatrices_SYK_SSF, Maldacena_2016_commentsonthesykmodel_ig:OTOC, PhysRevD.94.126010_spectralRandomMatrixSYK_levelstat} and Krylov complexity \cite{PhysRevX.9.041017_universalopgrowthhypothesis_parkerEtAl}, in both finite and infinite settings. Both the Coloured and Uncoloured Tensor Models exhibit the same large-N behaviour as the SYK Model, and thus possess the same behaviour in terms of OTOCs and Krylov complexity \cite{Carrozza:2015adg, 2016arXiv161009758W_edwardwitten, PhysRevD.95.046004_uncolored_model_into, PhysRevX.9.041017_universalopgrowthhypothesis_parkerEtAl}. Further, they have shown signs of chaos for finite $N$ \cite{Krishnan2017bala_colored, Krishnan2017_uncolored} in terms of the eigenvalue level-spacing distribution and the spectral form factor. The spectral properties of the $n = 3, d = 3$ Uncoloured Tensor Model were studied in \cite{Krishnan2017_uncolored}, which was the smallest non-trivial model to show signs of chaotic behaviour. In this work, we examine the $n = 3, d = 3$ Uncoloured Tensor Model in terms of Krylov complexity. 

The Uncoloured Tensor Model is constructed from a single fermionic field tensor (no colour distinction) with $D$ indices:
\begin{equation}
    \psi^{i_1i_2...i_D},
\end{equation}
where each index $i_a$ runs from $1\text{ to }n$ (and $n,D$ are independent). There are $N=n^D$ independent fermionic fields that obey the anti-commutation relation
\begin{equation}
    \{\psi^{i_1i_2...i_D},\psi^{j_1j_2...j_D}\} = \delta^{i_1j_1}\delta^{i_2j_2}...\,\delta^{i_Dj_D}.
\end{equation}\label{eq: fermionic_anticommutation_uncolored}
The model has an $O(n)$ symmetry associated to each index, whereby the symmetry group of the model is $G\sim O(n)^D$, which for $n = 3, d = 3$ model is $O(3)^3$. The Hamiltonian of the Uncoloured Tensor Model is given by
\begin{equation}
    H= \frac{J}{\sqrt{8}}\sum_{a_1,a_2}\sum_{b_1,b_2}\sum_{c_1,c_2} \psi^{a_1b_1c_1} \psi^{a_1b_2c_2} \psi^{a_2b_1c_2} \psi^{a_2b_2c_1}.
\end{equation}
In what follows, we fix $J=4$ for all numerical calculations. The fermions can be realised by gamma matrices. Since we have $N=27$, we work with gamma matrices of $SO(28)$ constructed in the chiral basis, following \cite{Krishnan2017_uncolored}. The gamma matrices are assigned to fermionic operators as
\begin{align}\label{eq: H uncolored n3d3}
  \sqrt{2}\,  \psi^{ijk}= \gamma_p, \ \ \text{where} \ \ 
    p = \, 9(i-1)+3(j-1)+k .
\end{align}
The Hamiltonian in this case is $16384$-dimensional, which is highly degenerate with only $34$ distinct eigenvalues\cite{Krishnan2017_uncolored} and is merely $0.39\%$ dense.

\subsection{Krylov Complexity}

\begin{figure}[t]
    \centering
    \begin{subfigure}[b]{0.48\textwidth}
        \centering
        \includegraphics[width=\linewidth]{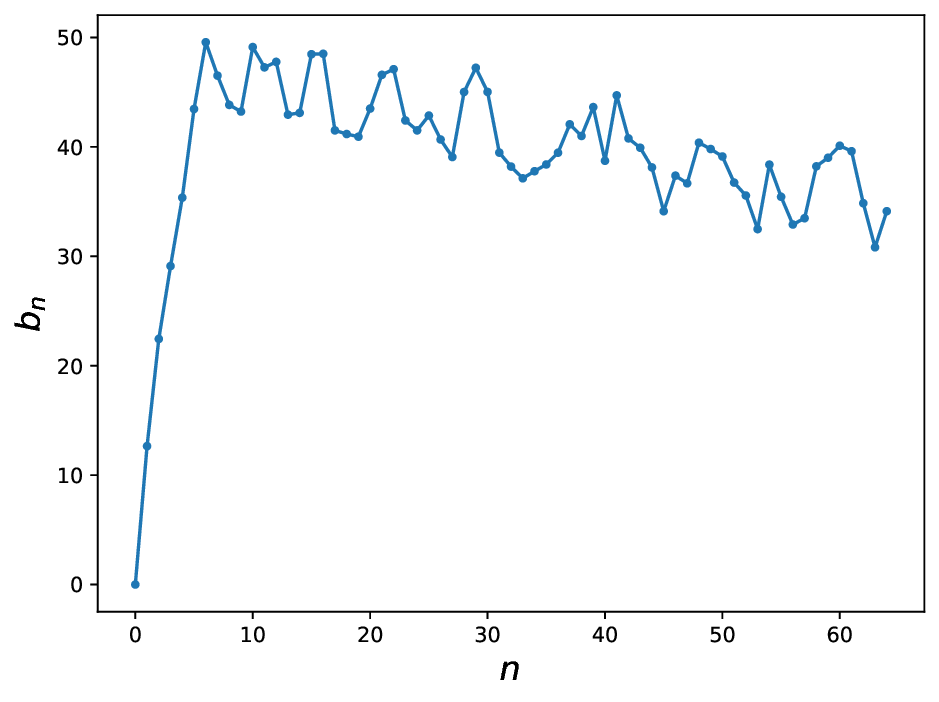}
        \caption{}
        \label{fig: Lanczos seq_inf temp}
    \end{subfigure}
    
    \vspace{0.1cm}
    
    \begin{subfigure}[b]{0.48\textwidth}
        \centering
        \includegraphics[width=\linewidth]{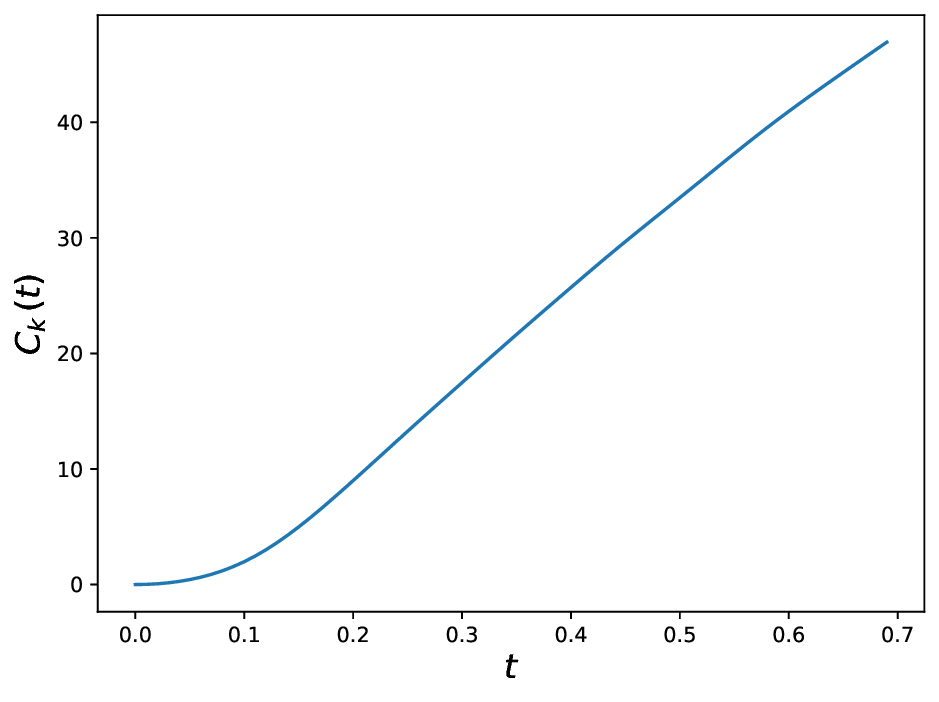}
        \caption{}
        \label{fig: Ck_0}
    \end{subfigure}
    \hfill
    \begin{subfigure}[b]{0.48\textwidth}
        \centering
        \includegraphics[width=\linewidth]{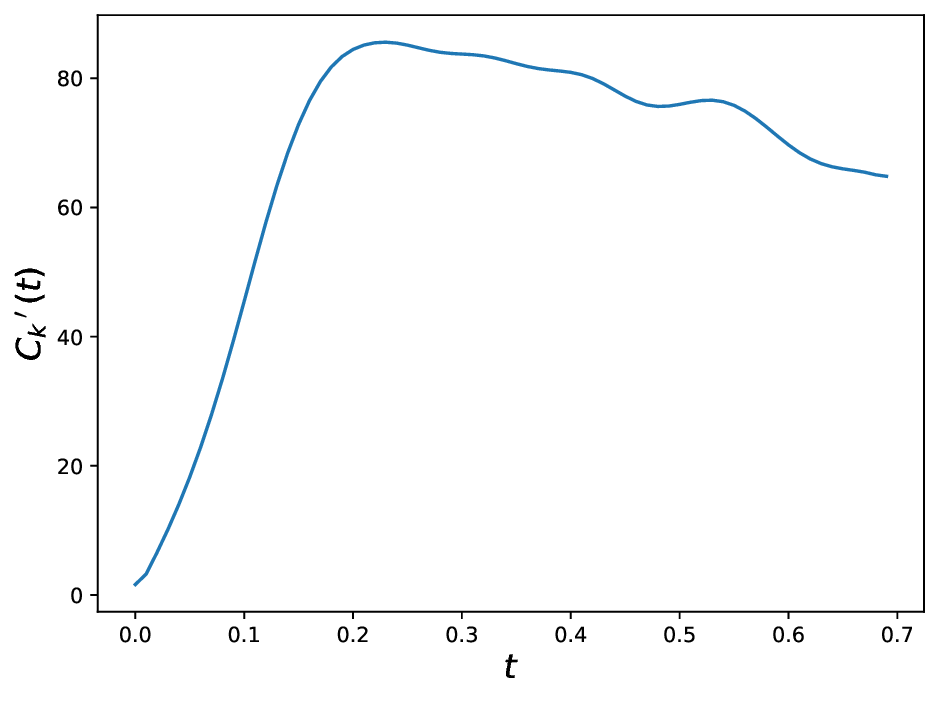}
        \caption{}
        \label{fig: Ck_0_derv}
    \end{subfigure}
    \caption{Lanczos coefficients (a), Krylov complexity (b) and its time derivative (c) of operator $\gamma_2$ in the infinite temperature limit.}
    \label{fig: Lanczos and Ck inf temp}
\end{figure}

In this section, we present the numerically evaluated Lanczos coefficients and Krylov complexity of the Uncoloured Tensor Model. The Lanczos coefficients $( b_n)$ were evaluated using the Lanczos algorithm with Partial Re-Orthogonalisation (PRO) (see \ref{subsec: FO PRO} for details). The Lanczos coefficients for a simple operator $\gamma_2=\sqrt{2}\,\psi^{112}$ in the infinite temperature limit are shown in Fig. \ref{fig: Lanczos seq_inf temp}. For finite-dimensional systems, the Krylov space dimension $K \leq D^2-D+1$, where $D$ is the dimension of the Hilbert space and $K$ is the number of eigenspaces of the Liouvillian over which the operator has a nonzero projection\cite{Rabinovici2021_Kcomp_thesis_I}. Here, $H$, despite being $16384$-dimensional, has numerous symmetries, and only $34$ distinct eigenvalues \cite{Krishnan2017_uncolored}. This translates to a high degeneracy in the spectrum of $\mathcal{L}$. For a general spectrum of $H$ with $34$ distinct eigenvalues, $\mathcal{L}$ is expected to have $1123$ distinct eigenvalues. But here, the presence of \textit{spectral mirror symmetry} about a mid value further decreases the number of unique eigenvalues of $\mathcal{L}$. We numerically find 537 unique eigenvalues for the Liouvillian. Additionally, we find that the Krylov space dimension $K$ for the operator $\gamma_2$ is 363 by explicitly computing the Liouvillian eigenspaces onto which the operator has a non-trivial projection. The Fig. \ref{fig: Lanczos seq_inf temp} shows the Lanczos sequence for $\gamma_2$ up to $n=65$ (terminated due to numerical instability of the Lanczos algorithm for degenerate matrices, see Appendix \ref{appndx: numerical_instability}).

In \cite{PhysRevX.9.041017_universalopgrowthhypothesis_parkerEtAl}, it is hypothesised that Lanczos coefficients in infinite-dimensional chaotic systems show asymptotic linear growth in both infinite and finite temperature limits. However, for systems with finite degrees of freedom, the linear growth of Lanczos coefficients is expected to last till $n\sim \log(K)$ for chaotic systems, followed by a plateau and decay \cite{Rabinovici2021_Kcomp_thesis_I}. In the Uncoloured Tensor Model, this initial growth phase lasts $\log(363)\sim 6 $ steps, as is observed in Fig. \ref{fig: Lanczos seq_inf temp}. The structure of the Lanczos coefficients we obtain is qualitatively very similar to that of the SYK Model with finite-$N$ obtained in \cite{Jian2021}; the major difference being that the plateau in the SYK case is very smooth, but the smallest model considered in \cite{Jian2021} has a $2^{10}$-dimensional Hilbert space, and $\sim 10^5$ Liouvillian eigenvalues. The truncated Lanczos sequence is sufficient to capture the small-time dynamics of Krylov complexity. Fig. \ref{fig: Ck_0} shows this early growth, evaluated numerically using the truncated Lanczos sequence, by solving equation (\ref{eq:phi_n Schroedinger eq}) for the wave function. However, the number of Lanczos coefficients available is insufficient to observe the expected saturation of Krylov complexity, as we encounter numerical instabilities much earlier.

\begin{figure}[tbp]
\centering
\begin{minipage}{0.48\textwidth}
\centering
\includegraphics[width=\linewidth]{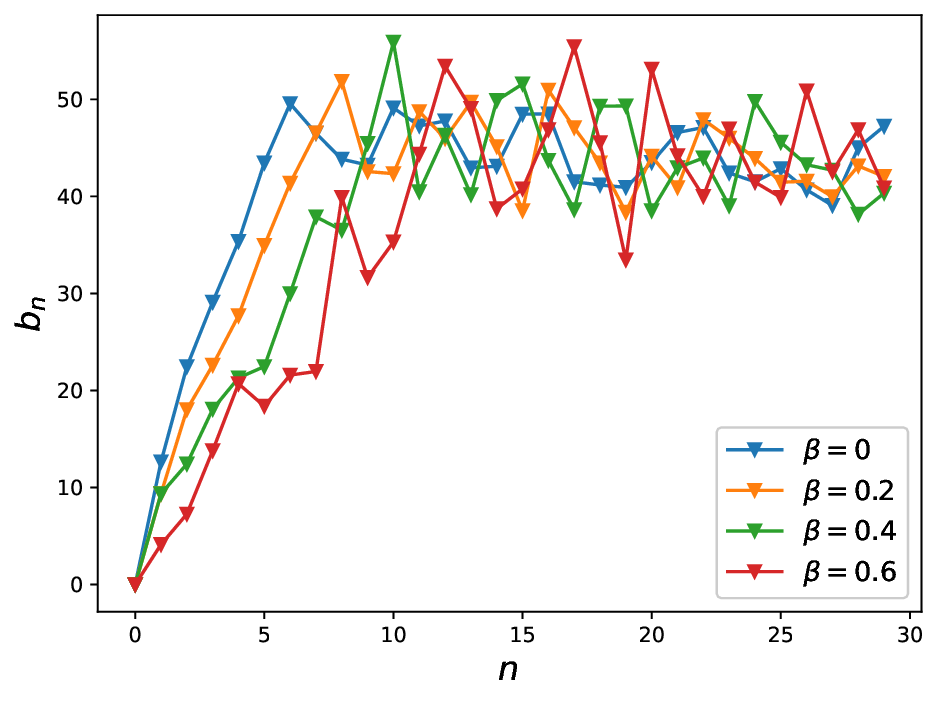}
\caption{\label{fig: Lanczos seq_finite temp all} Lanczos sequence $b_n$ of the operator $\sqrt{2} \psi^{112} =\gamma_2$, for different values of $\beta=\frac{1}{T}$.}
\end{minipage}
\hfill
\begin{minipage}{0.48\textwidth}
\centering
\includegraphics[width=\linewidth]{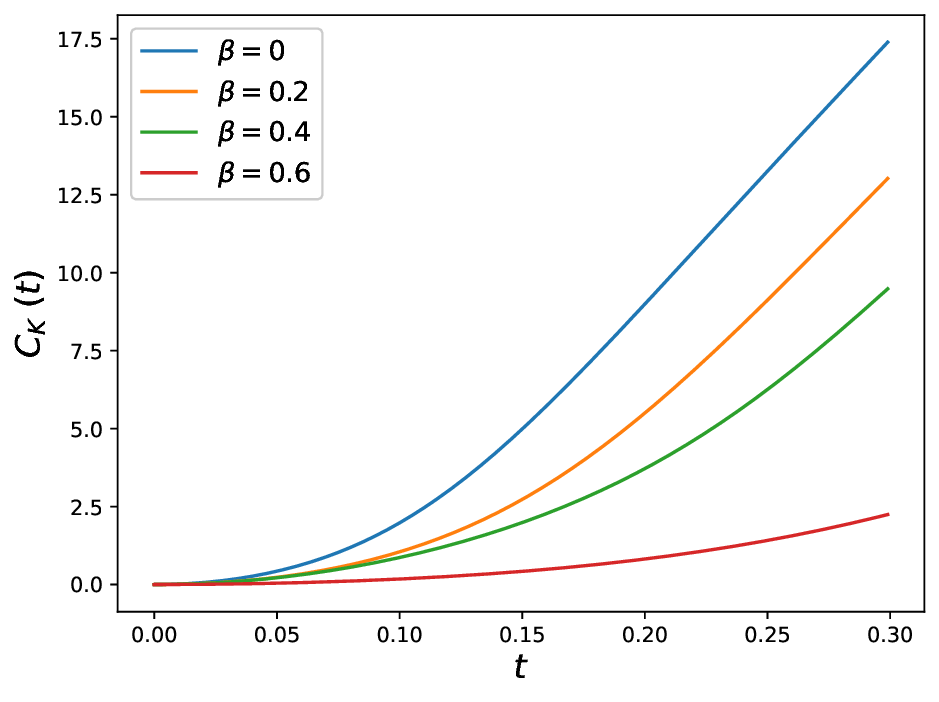}
\caption{\label{fig: K-comp_finite temp all} Krylov compelexity of the operator $\sqrt{2} \psi^{112} =\gamma_2$, for different values of $\beta=\frac{1}{T}$.}
\end{minipage}
\end{figure}

In the finite temperature limit, the initial growth of the Lanczos sequence is at a slower rate for lower temperature, see Fig. \ref{fig: Lanczos seq_finite temp all}. We find that the finite-temperature coefficients appear to reach the same plateau, but at larger values of $n$. The associated Krylov complexity at different temperatures is shown in Fig. \ref{fig: K-comp_finite temp all}, and we see that the growth of complexity is slower at lower temperatures. We have not computed the same for larger $\beta$ because the observed pattern is already consistent with similar systems, and numerical instability increases with increasing $\beta$.

\subsection{Symmetry-resolved Krylov Complexity}
The Uncoloured Tensor Model has numerous discrete symmetries and Noether charges owing to the global symmetry under the $O(n)^d$ rotations on the fermionic fields \cite{Krishnan2018_conrastingSYKlike, Klebanov_2018}. In the following, we discuss the symmetry-resolved Krylov complexity for the operator $\gamma_2 $, under a set of symmetries that commute with it.

\subsubsection*{Discrete Symmetries}

\begin{figure}[H]
\centering
\begin{minipage}{0.48\textwidth}
\centering
\includegraphics[width=\linewidth]{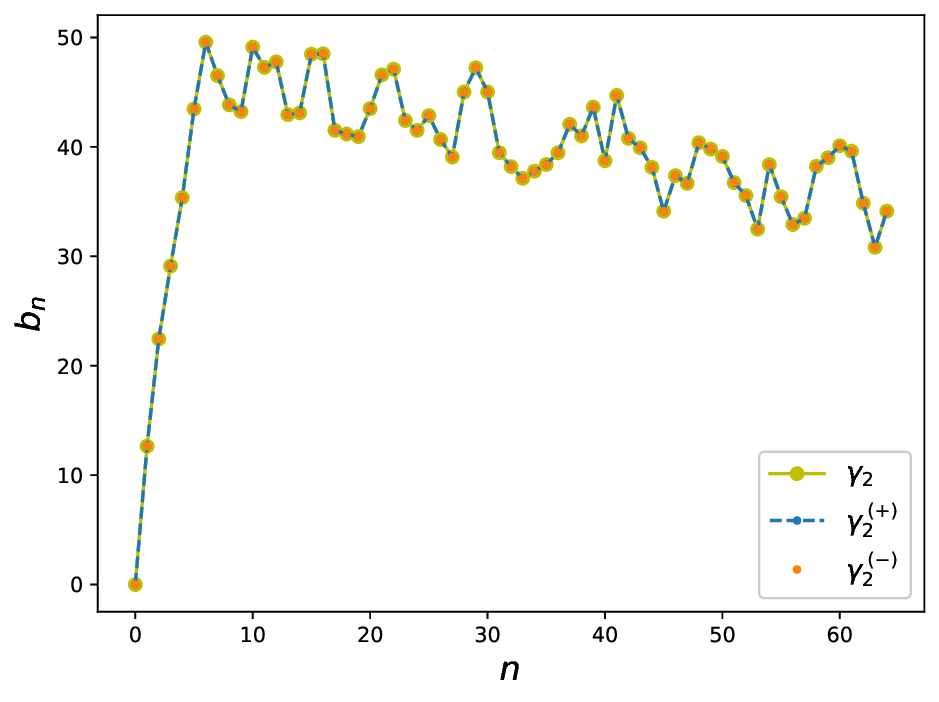}
\caption{\label{fig: Lanczos seq_symm_Z} Lanczos sequence $b_n$ for the $Z$-block operators of invariant operator $\gamma_2$.}
\end{minipage}
\hfill
\begin{minipage}{0.48\textwidth}
\centering
\includegraphics[width=\linewidth]{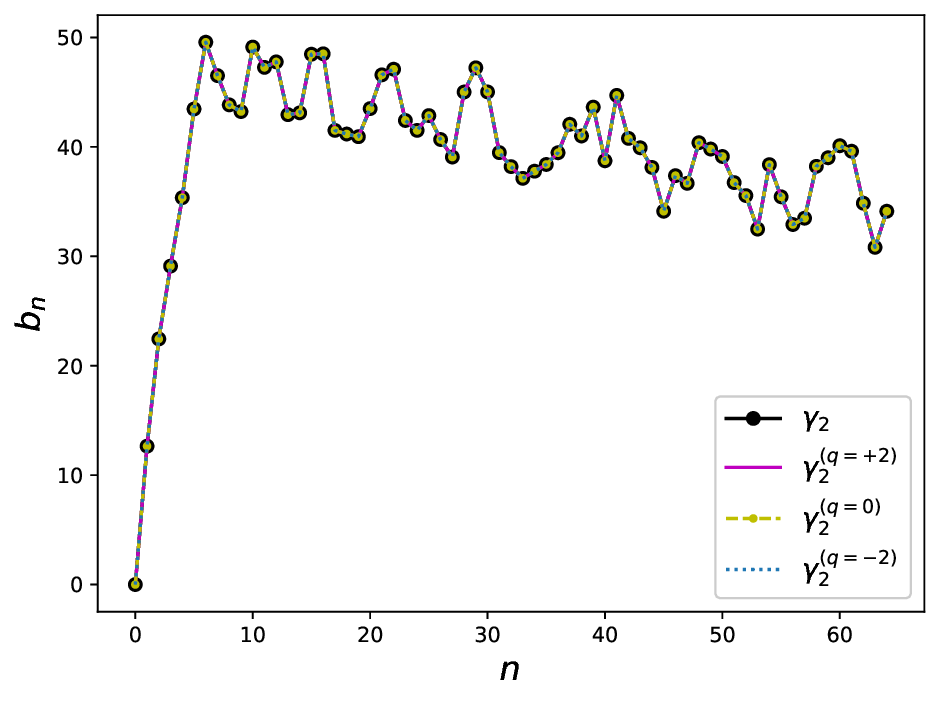}
\caption{\label{fig: Lanczos seq_symm_N1_Z_3spaces} Lanczos sequence $b_n$ for the $Q=N_1+Z$-block operators of invariant operator $\gamma_2$.}
\end{minipage}
\end{figure}

The $O(3)^3$ Uncoloured Tensor Model spectrum has a 16-fold degeneracy\cite{Krishnan2017_uncolored}, constituted by the discrete symmetries that can be written down in terms of products of the Gamma matrices. For brevity, we only discuss the operators we use for analysis (see Appendix \ref{appndx: symmetries of H} for a review of the symmetries). We consider two operators that commute with both the Hamiltonian and the operator $\gamma_2$,
\begin{align}
    Z &= (-i)\gamma_1\gamma_2\,...\gamma_{27},\\
    N_1 &= \gamma_1\gamma_2\,...\gamma_{9}.
\end{align}
Since both of these operators are products of gamma matrices, their eigenvalues are $\pm 1$. We can see that the unitary operator $A=PN_2$ satisfies the conditions (\ref{eq:cond_A1}) to (\ref{eq:cond_A3}), due to which the Hilbert space is \textit{identical} in both the $\pm 1$ sectors. Consequently, the condition (\ref{eq: gen_equipartition}) is automatically satisfied for all energy pairs, ensuring equipartition of Krylov complexity. We, nonetheless, numerically evaluate the Lanczos coefficients, shown in Fig. \ref{fig: Lanczos seq_symm_Z}, and confirm their precise matching up to $n=65$. 

Since $Z$ and $N_1$ commute, we can look for further equipartition among the $ +2$, $0$, and $-2$ blocks of the total charge $Z+N_1$ by finding operators that satisfy (\ref{eq:cond_B1})-(\ref{eq:cond_B3}). This is indeed achieved by $A= PN_1$ and $B=U$, which guarantees equipartition of the Krylov complexity among the three blocks. We also numerically verify the same, and Lanczos coefficients of the three blocks and the full operator are shown in Fig. \ref{fig: Lanczos seq_symm_N1_Z_3spaces}. The block $q=0$ is twice the size of the $q=\pm2$ blocks, as in effect it is made up of $Z=1, N_1=-1$ and $Z=-1, N_1=1$ blocks. Here, $B$ can be thought of as mapping the $N_1=\pm$ sectors at $Z=1$ and $-1$, resulting in four blocks with the same Lanczos coefficients. In other words, the $q=0$ block can be split into two distinct blocks which have the same Lanczos coefficients. 

\subsubsection*{Noether Symmetries}

\begin{figure}[H]
    \centering
    \begin{subfigure}[b]{0.48\textwidth}
            \centering
            \includegraphics[width=\linewidth]{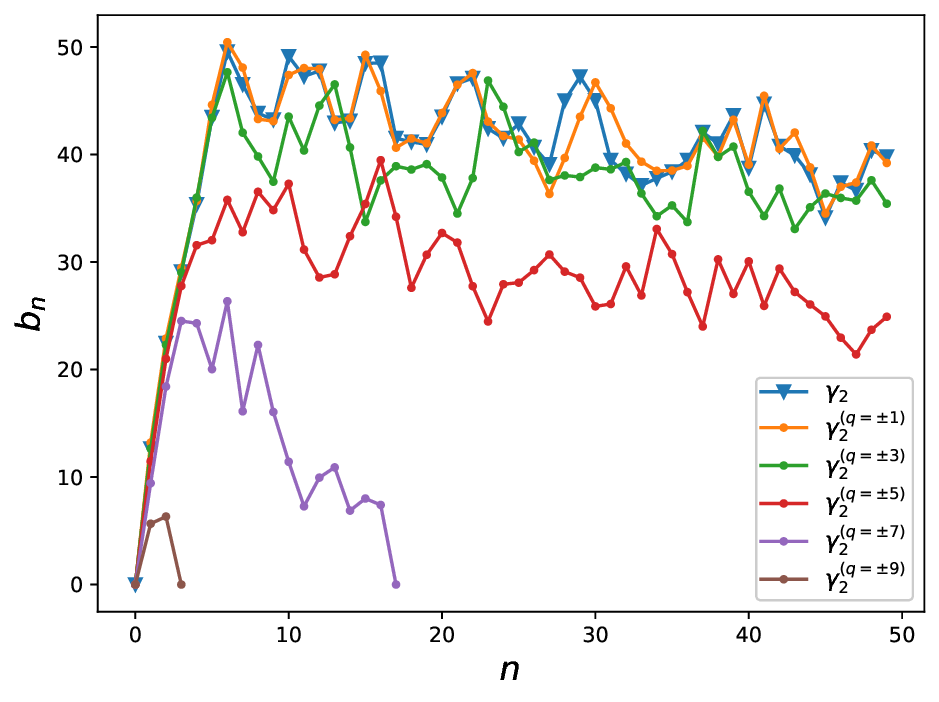}
            \caption{}
            \label{fig: Lanczos seq_symm_Q123_all}
    \end{subfigure}
    
    \vspace{0.1cm}
    
    \begin{subfigure}[b]{0.48\textwidth}
        \centering
        \includegraphics[width=\linewidth]{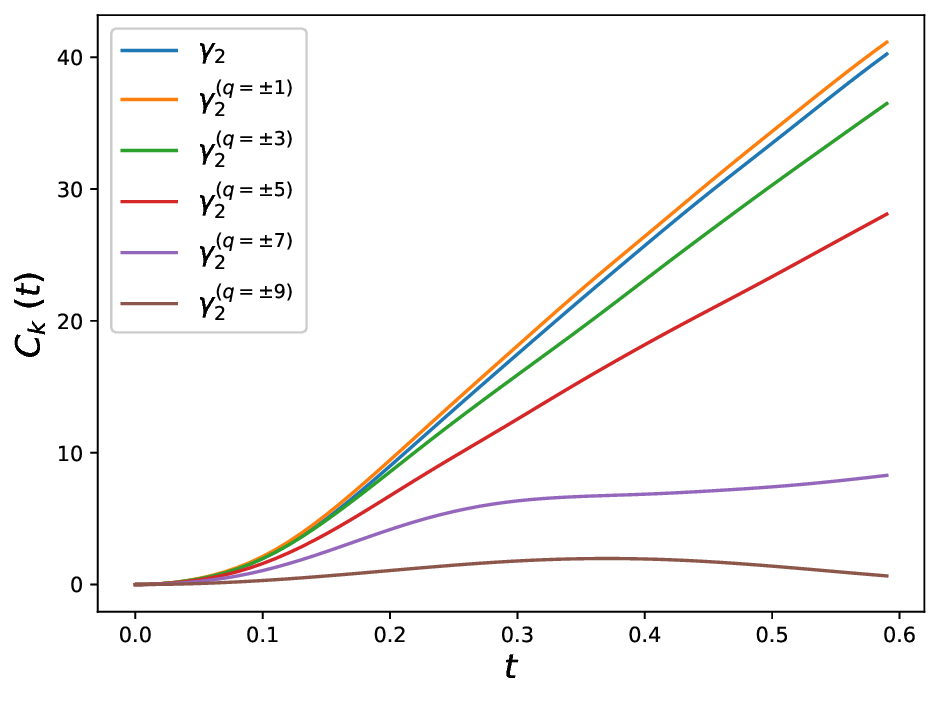}
        \caption{}
    \end{subfigure}
    \hfill
    \begin{subfigure}[b]{0.48\textwidth}
        \centering
        \includegraphics[width=\linewidth]{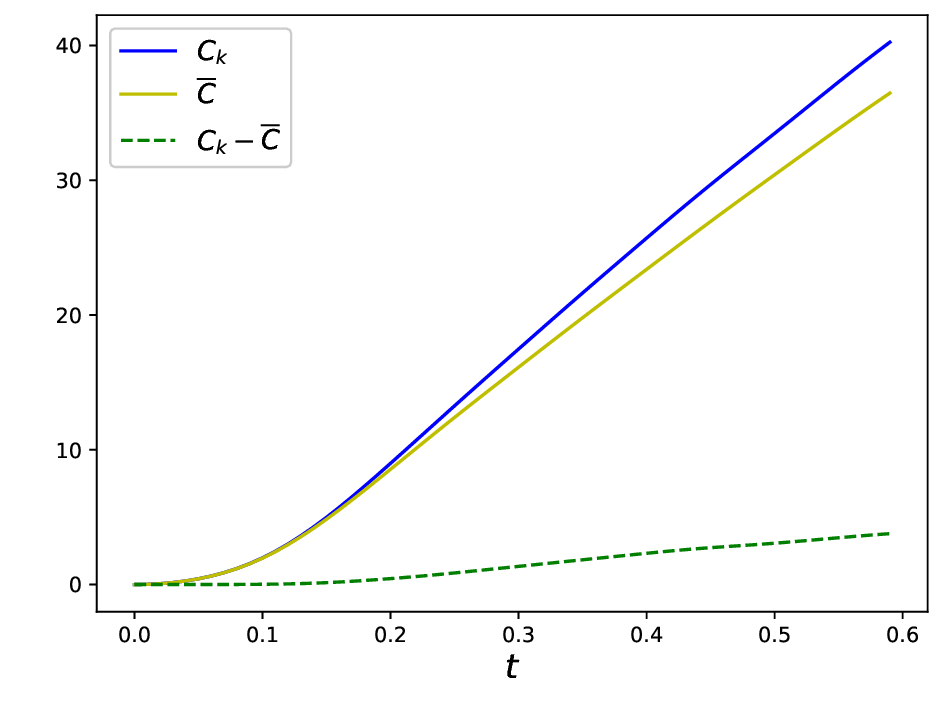}
        \caption{}
    \end{subfigure}
    \caption{Lanczos sequence $b_n$ (a) and Krylov complexity of the full operator and symmetry-resolved Krylov complexities of the block operators (b) of the invariant operator $\gamma_2$ in the eigenbasis of Noether charge $Q_1^{23}$. The growth of the average Krylov complexity across charge sectors, compared with the Krylov complexity of the full operator, is shown in (c). }
    \label{fig: Lanczos, Ck and Cbar of Q123}
\end{figure}
The Lagrangian of the Uncoloured Tensor Model is invariant under the continuous transformation of the fermionic fields
\begin{equation}
    \psi^{ijk} \rightarrow M_1^{ii'} M_2^{jj'} M_3^{kk'} \psi^{i'j'k'} \quad ; \,M_1,M_2,M_3 \in O(3).
\end{equation}
The corresponding Noether charges are
\begin{align}
    Q_{1}^{ab}&=i\,\psi^{ajk}\psi^{bjk},\\
    Q_{2}^{ab}&=i\,\psi^{iak}\psi^{ibk},\\
    Q_{3}^{ab}&=i\,\psi^{ija}\psi^{ijb},
\end{align}
where  $a,b\in\{1,2,3\}$ and $a\ne b$, which amounts to a total of $9$ conserved charges. In the quantum model, $H$ commutes with these $9$ Hermitian operators, which are the generators of the global $O(3)^3$ symmetry. Of this, we pick $Q_1^{23}$ as it commutes with $\gamma_2$. The charge subspaces in this case are interesting because the energy spectrum is not identical across them. In the largest charge sector $q=9$, the operator $\gamma_2$ has components along just three pairs of $(a,b)$, which is substantially different from that in the full space. We have explicitly checked the validity of (\ref{eq: gen_equipartition}) using the joint eigenbasis of $H$ and $Q$ obtained by numerical diagonalisation, for all pairs of $(a,b)$ in each charge subspace and we find the following: in the charge subspaces $q\neq \pm 1$, there are $(a,b)$ pairs over which the full operator has a non-trivial projection, whereas in the charge subspace, the operator has no projection over the same pair; in the charge subspace $q=\pm 1$, the pairs over which the operator has projection in the full space and the charge subspace are the same, however, the condition for equipartition is not satisfied in any of the non-trivial $(a,b)$ pairs.

Even though the individual charge sectors do not satisfy equipartition, we can see that the Lanczos coefficients of the blocks with opposite charges are identical. This can be easily verified as the symmetry operator $U$ satisfies the conditions in (\ref{eq:cond_A1})-(\ref{eq:cond_A3}), which leads to 
\begin{align}
    \text{Tr}(\Pi_{b,-q}\calo^\dagger\Pi_{a,-q}\calo)  = \text{Tr}(\Pi_{b,q}\calo^\dagger\Pi_{a,q}\calo).
\end{align}
We can also extend this to show
\begin{equation}
    \text{Tr}(\Pi_{b,-}\calo^\dagger\Pi_{a,-}\calo)  = \text{Tr}(\Pi_{b,+}\calo^\dagger\Pi_{a,+}\calo),
\end{equation}
which means we get equipartition between the positive and negative sectors of the Noether charge. Since there is an equipartition in the positive and negative charged sectors, we can look at the sectors within, say, the positive side and observe the following: to have equipartition in the $+1$ sector, the same has to apply in the complementary positive charge sector; this is impossible as the complementary sector does not have the full spectrum. Thus, the $+1$ sector, despite having all the energy eigenvalues of the full Hamiltonian, fails to exhibit equipartition. Additionally, as per our discussions on equipartition for Noether symmetries, an operator with projections onto the multiplicity space that are not identical across different irreducible representations can exhibit equipartition only if careful cancellations occur, which is very unlikely. Since we have different spectra at different $q$ values, it is not surprising that the equipartition does not hold for the operator $\gamma_2$.

\section{Concluding Remarks} \label{sec: conclusions}
In the presence of symmetries for a quantum system, it would be a significant simplification of computation if we could ascertain that the Krylov complexity of an operator in a charge subspace is identical to that in the full space. This, especially in the systems that are only accessible numerically, allows for investigations on larger systems where the presence of chaos is more discernible. We establish that, for equipartition of Krylov complexity to hold in the charge sectors, the condition (\ref{eq: gen_equipartition}) must be satisfied in all of them. However, to check the validity of the condition in any quantum system, the eigensystem must be known; this is quite computationally expensive if numerical diagonalisation is necessary. 

From further analysis, we also obtained the conditions (\ref{eq:cond_A1}-\ref{eq:cond_A3}) that guarantee equipartition in symmetries that have two charge sectors, and (\ref{eq:cond_B1}-\ref{eq:cond_B3}) for a combination of such symmetries. The existence of such operators can be studied using the symmetry algebra of the Hamiltonian, and if we can find them, the computation can be carried out in a charge subspace. For symmetries that have more than two charge sectors, a more general condition was obtained (\ref{eq: ij_matching}), which is only accessible if the form of the projection operators to the charge subspaces is already known. We also consider the case where the system has continuous symmetries and the operator commutes with at least one generator, keeping in mind the representation theory of $SU(2)$. We argue that equipartition holds for operators in a very restricted subset, depending on the structure of the Hilbert space in terms of irreducible representations. In all, we find that merely having the exact spectrum in any charge sector does not guarantee equipartition; the condition depends on the operator's norm in the appropriate subspaces. Also, it is worth noting that equipartition at any given temperature does not imply equipartition across all temperatures, as the conditions required for equipartition across all temperatures are far more restrictive.

In Section \ref{sec: tensor model}, we study the Krylov complexity of the $n=3,d=3$ Uncoloured Tensor Model numerically. We note that the Lanczos coefficients of the operator $\psi^{112}$ grow linearly for $6 \sim \log(363)$ steps, where 363 is the size of the Krylov dimension of the operator. This is followed by a plateauing of the Lanczos coefficients, and we were unable to carry out the numerics beyond $65$ steps because the errors became dominant. Nonetheless, we were able to study the early non-linear growth, followed by linear growth, of the Krylov complexity. The $n=3,d=3$ model is the smallest non-trivial model that is numerically accessible, with only $34$ distinct energy eigenvalues. This, at best, leads to a maximum Krylov Dimension of 1123 and an initial growth phase for up to $\log(1123)\sim 7$ steps. We find that the Lanczos coefficients obtained in this case are similar to those reported for SYK Models at small $N$, consistently placing the Tensor Model in line with SYK.

We also use the myriad of symmetries of the Uncoloured Tensor Model to validate and provide a contextual example to the conditions obtained in Section \ref{sec: conditions} for the existence of equipartition in charge subspaces. We show that operators that ensure equipartition are identifiable in the model and further confirm numerically that the Lanczos coefficients obtained in the full space and in each charge subspace are identical. We also study the Lanczos coefficients in the charge subspaces of the Noether symmetries and find that equipartition is non-existent, and for very justifiable reasons based on the structure of the operator. 

Looking ahead, the conditions we have obtained were for finite-dimensional systems, and it will be useful, particularly for applications to field-theoretic systems, to extend the analysis to infinite-dimensional systems. Also, similar conditions could be established for the spread complexity of the states, in both finite- and infinite-dimensional settings. Finally, although we find some constructive approaches that guarantee equipartition, especially in cases with twofold symmetry, it would be useful to work out similar strategies for other non-trivial groups. We hope to return to these questions in subsequent works.

\section*{Acknowledgements}
The authors thank Mr. Sreekanth S for useful discussions and collaboration during the early stage of the work. PNB acknowledges the financial support of the National Institute of Technology under the FRG Scheme FRG/2022/PHY\_01.

\appendix

\section{Conditions for equipartition: Detailed proof}\label{appndx: proof equipartition}
If the symmetry-resolved Krylov complexity in the $q_0$ subspace has to be identical to that of the full operator, which in turn requires $b_n = b_n^{(q_0)}$, the condition 
\begin{align}
    \dfrac{\norm{\call^n\calo}^2_\beta}{\norm{\calo}^2_\beta} = \dfrac{\norm{\call^n\calo_{q_0}}^2_\beta}{\norm{\calo_{q_0}}^2_\beta}
\end{align}
has to be satisfied for all $n$. Written in terms of the coefficients of $\calo $ expanded in the Liouvillian eigenbasis, this translates to 
\begin{align}
    \dfrac{\sum\limits_{a,b=1}^{N} \sum\limits_{q'=1}^{|Q|} \sum\limits_{\rho= 1}^{|\cala_{(a,q')}|} \sum\limits_{\sigma= 1}^{|\cala_{(b,q')}|} e^{-\frac{\beta}{2}(E_a+E_b)}\omega_{ab}^{2n}\, |c_{a,b,q'}^{\rho,\sigma}|^2}{\sum\limits_{a,b=1}^{N} \sum\limits_{q'=1}^{|Q|} \sum\limits_{\rho= 1}^{|\cala_{(a,q')}|} \sum\limits_{\sigma= 1}^{|\cala_{(b,q')}|}e^{-\frac{\beta}{2}(E_a+E_b)}\, |c_{a,b,q'}^{\rho,\sigma}|^2} =     \dfrac{\sum\limits_{a,b=1}^{N} \sum\limits_{\rho= 1}^{|\cala_{(a,q_0)}|} \sum\limits_{\sigma= 1}^{|\cala_{(b,q_0)}|}e^{-\frac{\beta}{2}(E_a+E_b)} \omega_{ab}^{2n}\, |c_{a,b,q_0}^{\rho,\sigma}|^2}{\sum\limits_{a,b=1}^{N} \sum\limits_{\rho= 1}^{|\cala_{(a,q_0)}|} \sum\limits_{\sigma= 1}^{|\cala_{(b,q_0)}|}e^{-\frac{\beta}{2}(E_a+E_b)}\, |c_{a,b,q_0}^{\rho,\sigma}|^2}.
\end{align}
To make the computation more transparent, let us introduce the notation, 
\begin{align}
    \sum\limits_{\rho= 1}^{|\cala_{(a,q)}|} \sum\limits_{\sigma= 1}^{|\cala_{(b,q)}|} |c_{a,b,q}^{\rho,\sigma}|^2 = l^2_{(a,b),q},
\end{align}
by means of which the preceding equation can be rewritten in the simplified form
\begin{align}\label{eq: cond-chi,l}
    \dfrac{\sum\limits_{a,b=1}^{N} \sum\limits_{q'=1}^{|Q|}  e^{-\frac{\beta}{2}(E_a+E_b)}\omega_{ab}^{2n}\, l^2_{(a,b),q'}}{\sum\limits_{a,b=1}^{N} \sum\limits_{q'=1}^{|Q|}e^{-\frac{\beta}{2}(E_a+E_b)}\, l^2_{(a,b),q'}} =    \dfrac{\sum\limits_{a,b=1}^{N} e^{-\frac{\beta}{2}(E_a+E_b)}\omega_{ab}^{2n}\,l^2_{(a,b),q_0}}{\sum\limits_{a,b=1}^{N}e^{-\frac{\beta}{2}(E_a+E_b)}\, l^2_{(a,b),q_0}}.
\end{align}
For a general Hamiltonian spectrum, it is possible that multiple pairs $(a,b)$ of energy eigenvalues lead to the same Liouvillian eigenvalue. The set of Liouvillian eigenvalues is given by $ \{\omega_\Lambda | \, \Lambda\in \{0,\pm1,\dots,\pm L \} \} $, which can be used to re-group the double-summation over the energy values as 
\begin{align}
    \sum_{a,b=1}^{N} (\dots) = \sum_{\Lambda=-L}^{L} \,\,\sum_{\substack{a,b \text{ s.t. } \\ \omega_{ab}=\omega_\Lambda}} (\dots) \equiv  \sum_{\Lambda=-L}^{L} \,\,\sum_{(a,b)|\Lambda} (\dots).
\end{align}
Thus, the condition in (\ref{eq: cond-chi,l}) can be simplified to 
\begin{align}
\sum_{\Lambda=-L}^{L} \omega_\Lambda^{2n}\, \psi_{\Lambda,q_0} &= 0,\\
\text{where    } \psi_{\Lambda,q_0} &= \sum_{(a,b)|\Lambda} e^{-\frac{\beta}{2}(E_a+E_b)}\left( \dfrac{ \sum\limits_{q'=1}^{|Q|}  \, l^2_{(a,b),q'}}{\sum\limits_{c,d=1}^{N} \sum\limits_{q'=1}^{|Q|}e^{-\frac{\beta}{2}(E_c+E_d)}\, l^2_{(c,d),q'}} -    \dfrac{ l^2_{(a,b),q_0}}{\sum\limits_{c,d=1}^{N}e^{-\frac{\beta}{2}(E_c+E_d)}\, l^2_{(c,d),q_0}}\right) \nonumber.
\end{align}
If the operator under consideration is Hermitian, the coefficients satisfy $c_{(a, b),q}^{(\rho,\sigma)} = (c_{(b, a),q}^{(\sigma,\rho)})^*  $, which subsequently gives the conditions $l^2_{(a,b),q} = l^2_{(b, a),q}$ and $\psi_{\Lambda,q} = \psi_{-\Lambda,q} $. The Liouvillian eigenvalues are symmetric about zero, whereby $\omega_\Lambda = - \omega_{-\Lambda} $ and $\omega_0 = 0$. Putting these together, we get $2L+1$ equations 
\begin{align}
     \delta_{0,n} \psi_{0,q_0} + 2\sum\limits_{\Lambda=1}^{L} \omega_\Lambda^{2n} \psi_{\Lambda,q_0}  =0, \ \ \ \text{where } n\in 0,1,\dots,2L.
\end{align}
Since $\omega_\Lambda$ are distinct, the unique solution to the set of equations is $\psi_{\Lambda,q_0} = 0 $ for all values of $\Lambda$, which means
\begin{align}
    \dfrac{\sum\limits_{(a,b)|\Lambda} e^{-\frac{\beta}{2}(E_a+E_b)} \sum\limits_{q'=1}^{|Q|}  \, l^2_{(a,b),q'}}{\sum\limits_{c,d=1}^{N} \sum\limits_{q'=1}^{|Q|}e^{-\frac{\beta}{2}(E_c+E_d)}\, l^2_{(c,d),q'}} &=    \dfrac{\sum\limits_{(a,b)|\Lambda} e^{-\frac{\beta}{2}(E_a+E_b)}  l^2_{(a,b),q_0}}{\sum\limits_{c,d=1}^{N}e^{-\frac{\beta}{2}(E_c+E_d)}\, l^2_{(c,d),q_0}} \\
    \Rightarrow \ \  \dfrac{\sum\limits_{(a,b)|\Lambda} e^{-\frac{\beta}{2}(E_a+E_b)}  l^2_{(a,b),q_0}}{\sum\limits_{(a,b)|\Lambda} e^{-\frac{\beta}{2}(E_a+E_b)} \sum\limits_{q'=1}^{|Q|}  \, l^2_{(a,b),q'}}  &=    \dfrac{\sum\limits_{c,d=1}^{N}e^{-\frac{\beta}{2}(E_c+E_d)}\, l^2_{(c,d),q_0}}{\sum\limits_{c,d=1}^{N} \sum\limits_{q'=1}^{|Q|}e^{-\frac{\beta}{2}(E_c+E_d)}\, l^2_{(c,d),q'}} \ \ \forall \Lambda . 
\end{align}
Crucially, we can pick any two Liouvillian eigenspaces $\Lambda$ and $\Lambda'$, and obtain
\begin{align}
     \sum\limits_{(a,b)|\Lambda} \sum\limits_{(c,d)|\Lambda'}  e^{-\frac{\beta}{2}(E_a+E_b+E_c+E_d)} l^2_{(a,b)}l^2_{(c,d)} \left[ \dfrac{l^2_{(a,b),q_0}}{l^2_{(a,b)}} - \dfrac{l^2_{(c,d),q_0}}{l^2_{(c,d)}} \right] = 0,
\end{align}
where $ \sum\limits_{q'=1}^{|Q|} l^2_{(a,b),q'} = l^2_{(a,b)} $. The condition can be trivially met if all terms in the square brackets vanish for all pairs of energies. However, at any given temperature, this is only a sufficient condition and not a necessary one. An example that elucidates this is given at the end of this appendix. If we demand that equipartition holds for all temperatures, we can show that it becomes a necessary condition as well. To see this, let us rewrite the condition as
\begin{align}
    \sum\limits_X e^{-\frac{\beta}{2}X} \left[\sum\limits_{ \substack{(a,b)|\Lambda \ ,\ (c,d)|\Lambda' \\ E_a+E_b+E_c+E_d= X}}l^2_{(a,b)}l^2_{(c,d)} \left( \dfrac{l^2_{(a,b),q_0}}{l^2_{(a,b)}} - \dfrac{l^2_{(c,d),q_0}}{l^2_{(c,d)}} \right)  \right] =0 .
\end{align}
If this condition is valid for all $\beta$, then it demands that the term in the square bracket is zero, for all values of $X$. In the Liouvillian eigenspace, the largest eigenvalue is formed of a single energy pair $\omega_L = E_{max} - E_{min}$ and in any particular eigenspace, the sum of energies of any pair is unique. Thus, we choose $\Lambda' = L $ and iterate over all values of $\Lambda$ so that any value of $X$ is given by a unique set of energies $(E_a,E_b)$ and $(E_{max},E_{min})$, and consequently we get
\begin{align}
    \dfrac{l^2_{(a,b),q_0}}{l^2_{(a,b)}} = \dfrac{l^2_{(max,min),q_0}}{l^2_{(max,min)}},
\end{align}
for all pairs of energies. We therefore have a very strict condition that ensures equipartition, given by
\begin{align}
    \dfrac{l^2_{(a,b),q_0}}{l^2_{(a,b)}} = \dfrac{l^2_{(c,d),q_0}}{l^2_{(c,d)}} = \eta_{q_0}, \ \ \forall (a,b)\ \text{and} \ (c,d),
\end{align}
where $\eta_{q_0}$ is a constant. Written in terms of the coefficients, this leads to
\begin{align}\label{eq:qmatching_ap}
    \dfrac{\sum\limits_{\rho = 1}^{|\cala_{(a,q_0)}|} \sum\limits_{\sigma = 1}^{|\cala_{(b,q_0)}|} |c_{a,b,q_0}^{\rho,\sigma}|^2} {\sum\limits_{q'=1}^{|Q|}\sum\limits_{\rho = 1}^{|\cala_{(a,q')}|} \sum\limits_{\sigma = 1}^{|\cala_{(b,q')}|} |c_{a,b,q'}^{\rho,\sigma}|^2} = \eta_{q_0}  =  \dfrac{\sum\limits_{a,b=1}^{N}\sum\limits_{\rho = 1}^{|\cala_{(a,q_0)}|} \sum\limits_{\sigma = 1}^{|\cala_{(b,q_0)}|} |c_{a,b,q_0}^{\rho,\sigma}|^2}{\sum\limits_{a,b=1}^{N}\sum\limits_{q'=1}^{|Q|}\sum\limits_{\rho = 1}^{|\cala_{(a,q')}|} \sum\limits_{\sigma = 1}^{|\cala_{(b,q')}|} |c_{a,b,q'}^{\rho,\sigma}|^2}.
\end{align}
In the complementary subspace of $q_0$, labelled by $\overline{q_0}$, we have 
\begin{align}
    \dfrac{\sum\limits_{q'\in\, \overline{q_0}}\sum\limits_{\rho = 1}^{|\cala_{(a,q_0)}|} \sum\limits_{\sigma = 1}^{|\cala_{(b,q_0)}|} |c_{a,b,q'}^{\rho,\sigma}|^2} {\sum\limits_{q'=1}^{|Q|}\sum\limits_{\rho = 1}^{|\cala_{(a,q')}|} \sum\limits_{\sigma = 1}^{|\cala_{(b,q')}|} |c_{a,b,q'}^{\rho,\sigma}|^2} = 1- \eta_{q_0} ,
\end{align}
which results in the Lanczos coefficients being the same in the $\overline{q_0}$ subspace as well, i.e. $b_n = b_n^{(\overline{q_0})}$. We could also rearrange (\ref{eq:qmatching_ap}) as
\begin{align}\label{eq:abmatching_ap}
    \dfrac{\sum\limits_{\rho = 1}^{|\cala_{(a,q_0)}|} \sum\limits_{\sigma = 1}^{|\cala_{(b,q_0)}|} |c_{a,b,q_0}^{\rho,\sigma}|^2} {\sum\limits_{a,b=1}^{N}\sum\limits_{\rho = 1}^{|\cala_{(a,q_0)}|} \sum\limits_{\sigma = 1}^{|\cala_{(b,q_0)}|} |c_{a,b,q_0}^{\rho,\sigma}|^2} = \dfrac{\sum\limits_{q'=1}^{|Q|}\sum\limits_{\rho = 1}^{|\cala_{(a,q')}|} \sum\limits_{\sigma = 1}^{|\cala_{(b,q')}|} |c_{a,b,q'}^{\rho,\sigma}|^2}{{\sum\limits_{a,b=1}^{N}\sum\limits_{q'=1}^{|Q|}\sum\limits_{\rho = 1}^{|\cala_{(a,q')}|} \sum\limits_{\sigma = 1}^{|\cala_{(b,q')}|} |c_{a,b,q'}^{\rho,\sigma}|^2}}, 
\end{align}
such that the right-hand side has no dependence on $q_0$ whatsoever. However, both sides depend explicitly on $(a,b)$, and we can expect that the ratio could, in general, depend on $(a,b)$, whereby
\begin{align}\label{eq:abmtching_z}
    \dfrac{\sum\limits_{\rho = 1}^{|\cala_{(a,q_0)}|} \sum\limits_{\sigma = 1}^{|\cala_{(b,q_0)}|} |c_{a,b,q_0}^{\rho,\sigma}|^2} {\sum\limits_{a,b=1}^{N}\sum\limits_{\rho = 1}^{|\cala_{(a,q_0)}|} \sum\limits_{\sigma = 1}^{|\cala_{(b,q_0)}|} |c_{a,b,q_0}^{\rho,\sigma}|^2} = \zeta_{a,b} , \ \text{such that  } \sum\limits_{a,b=1}^{N} \zeta_{a,b} = 1.
\end{align}
From the equations (\ref{eq:qmatching_ap}) and (\ref{eq:abmtching_z}), we find the condition on the coefficients
\begin{align}
        \dfrac{\sum\limits_{\rho = 1}^{|\cala_{(a,q_0)}|} \sum\limits_{\sigma = 1}^{|\cala_{(b,q_0)}|} |c_{a,b,q_0}^{\rho,\sigma}|^2}{\sum\limits_{a,b=1}^{N}\,\sum\limits_{q'=1}^{|Q|}\sum\limits_{\rho = 1}^{|\cala_{(a,q')}|} \sum\limits_{\sigma = 1}^{|\cala_{(b,q')}|} |c_{a,b,q'}^{\rho,\sigma}|^2} &= \eta_{q_0}\, \zeta_{a,b}\,,
\end{align}
in all $(a,b,q_0)$ subspaces, for an operator that admits equipartition. Introducing the projection operators,
\begin{align}
    \Pi_{a,q} =  \sum_\rho \ket{E_a,q,\rho}\bra{E_a,q,\rho}, \ \
    \Pi_{q} = \sum_a \Pi_{a,q}, \ \ \Pi_a = \sum_q \Pi_{a,q},
\end{align}
the condition in (\ref{eq:qmatching_ap}) can be written in a basis independent manner as 
\begin{align}
    \dfrac{\text{Tr}\left(\Pi_{b,q_0} \calo^\dagger \Pi_{a,q_0} \calo  \right)}{\text{Tr}\left(\Pi_{q_0} \calo^\dagger \calo  \right)} =    \dfrac{\text{Tr}\left(\Pi_{b} \calo^\dagger \Pi_{a} \calo  \right)}{\text{Tr}\left(\calo^\dagger \calo  \right)} .
\end{align}
\subsection{An Illustrative Example on Equipartition at a single temperature}
Consider a two-level system with energies labelled by $e, e'$ and two charge sectors labelled by $q, q'$. There are three Liouvillian eigenstates $\omega_1 = E_e - E_{e'} $, $\omega_0 = 0 = E_e-E_e = E_{e'} - E_{e'}$ and $\omega_{-1} = e'-e$. Further, let us consider an operator with the following subspace norms
\begin{align}
    l^2_{(e,e),q} = 1,  \ \ l^2_{(e',e'),q} = 9, \ \ l^2_{(e,e'),q} = 7,\ \
l^2_{(e,e),q'} = 7,  \ \  l^2_{(e',e'),q'} = 3, \ \  l^2_{(e,e'),q'} = 7.
\end{align}
The only non-trivial check to be performed is when $\Lambda \neq \Lambda'$, so let us work with $\Lambda = 1 $ and $\Lambda' = 0 $. In the $q$-sector and at arbitrary temperature, we get two terms with distinct total energies
\begin{align}
   &e^{-\frac{\beta}{2}(3E_e+E_{e'})}\left( l^2_{(e,e'),q} \, l^2_{(e,e)} - l^2_{(e,e),q} \, l^2_{(e,e')} \right) +  e^{-\frac{\beta}{2}(E_e+3E_{e'})}\left( 
    l^2_{(e,e'),q} \, l^2_{(e',e')} - l^2_{(e',e'),q}\, l^2_{(e,e')} \right)  \nonumber \\
    &= 42 \left( e^{-\frac{\beta}{2}(3E_e+E_{e'})} -  e^{-\frac{\beta}{2}(E_e+3E_{e'})} \right) .
\end{align}
This equation becomes zero only at infinite temperature ($\beta = 0$). Thus, for the operator we have considered, equipartition will hold only at infinite temperature. 

\section{Numerical Instability of Lanczos algorithm}\label{appndx: numerical_instability}
\subsection{Lanczos Algorithm} \label{sec: numerical details}
The Lanczos algorithm or recursion method is an iterative method historically developed to solve the numerical eigenvalue problem \cite{viswanath_2025_avfvz-v0s16}.
By the iterative action of the matrix (to be diagonalised) on some probe vector, the algorithm generates an orthogonalised set of vectors, called the \textit{Krylov Basis}, in which the matrix is tridiagonal.  

In a quantum system with Hilbert space $\mathcal{H}$, the set of all operators in $\mathcal{H}$ itself forms a vector space. The  Liouvillian $\call=[H, \cdot \,]$ can be thought of as a (super-)operator in the \textit{operator-vector space}. Let $\cbraket{}{}$ be an inner product in the operator space and $\cket{\calo}$ be the operator of interest ($\cket{\calo}$  assumed to be normalised). The Lanczos algorithm to generate the  $K$-dimensional Krylov basis $\{\cket{K_n}\}$ is as follows:
\begin{equation}
\label{eq:Lanczos alg first}
\begin{aligned}
    n=0 : \quad & \cket {K_0} = \cket{\calo}, \quad b_0=0,\\
    n\geq 1 : \quad & \cket{A_n} = \call\cket{K_{n-1}} - b_{n-1}\cket{K_{n-2}} \\
    & b_n = \cbraket{A_n }{A_n }^{1/2} \\
    & \text{if } b_n \leq \epsilon: \text{STOP, else: } \cket{K_n} = \frac{\cket{A_n}}{b_n} 
\end{aligned}
\end{equation}

Here, $b_n$ is called the \textit{Lanczos coefficient}.
Theoretically, the algorithm terminates when $b_n=0$, which may not occur in finite-precision arithmetic; hence, a tolerance value $\epsilon$ is set.

$\call$ is tri-diagonal in the Krylov basis,
\begin{equation}
    \cbra{K_n}\call\cket{K_m } = b_{m+1} \delta_{n,m+1} + b_m \delta _{n,m-1} + a_n \delta_{n,m}\,,
\end{equation}

\begin{equation}
\call_{nm}=
    \begin{bmatrix}
    0 & b_1 & 0 & \cdots & 0 \\
    b_1 & 0 & b_2 & \ddots & \vdots \\
    0 & b_2 & 0 & \ddots & 0 \\
    \vdots & \ddots & \ddots & \ddots & b_{n-1} \\
    0 & \cdots & 0 & b_{n-1} & 0
    \end{bmatrix}
\end{equation}

The Lanczos algorithm produces only one Krylov vector corresponding to each eigenspace of $\call$ over which $\cket{\calo}$ has a non-trivial projection. In the case where $\cket{\calo}$ has a non-trivial projection over all $\call$ eigenspaces, the spectrum of the $K\times K$ dimensional $\call_{nm}$ matrix has the same eigenvalues as $\call$, but non-degenerate.

\subsubsection{FO and PRO}\label{subsec: FO PRO}

The Lanczos algorithm fails to generate orthonormal vectors when working with finite precision \cite{Paige1971_thesis_errorLanczos}. The error accumulates with each iteration, so after a few iterations, the new vector is no longer orthogonal to the previous ones. Two alternate algorithms used to maintain orthogonality are: Full Orthogonalisation (FO) and Partial Re-Orthogonalisation (PRO) \cite{Simon1984_Lanczos_PRO}. In FO, in each iteration $n$, after evaluating the $n^{th}$ vector $\cket{K_n}$, it is explicitly orthogonalised with respect to all previous vectors:
\begin{align}
    &\cket{K_m}=\cket{K_m} -\cbraket{K_n}{K_m} \cket{K_n}  &\textit{for all:  }\,\,m<n
\end{align}
This is computationally expensive when dealing with large dimensions. In PRO, explicit re-orthogonalisation is not done in each iteration. Instead, an estimate of error build-up is calculated in parallel with the computations. When the error estimate reaches a set tolerance, the last two vectors are explicitly orthogonalised with respect to all previous vectors.

\subsection{Numerical Instability in Degenerate Operators}

\begin{figure}[ht]
\centering
\includegraphics[width=0.48\linewidth]{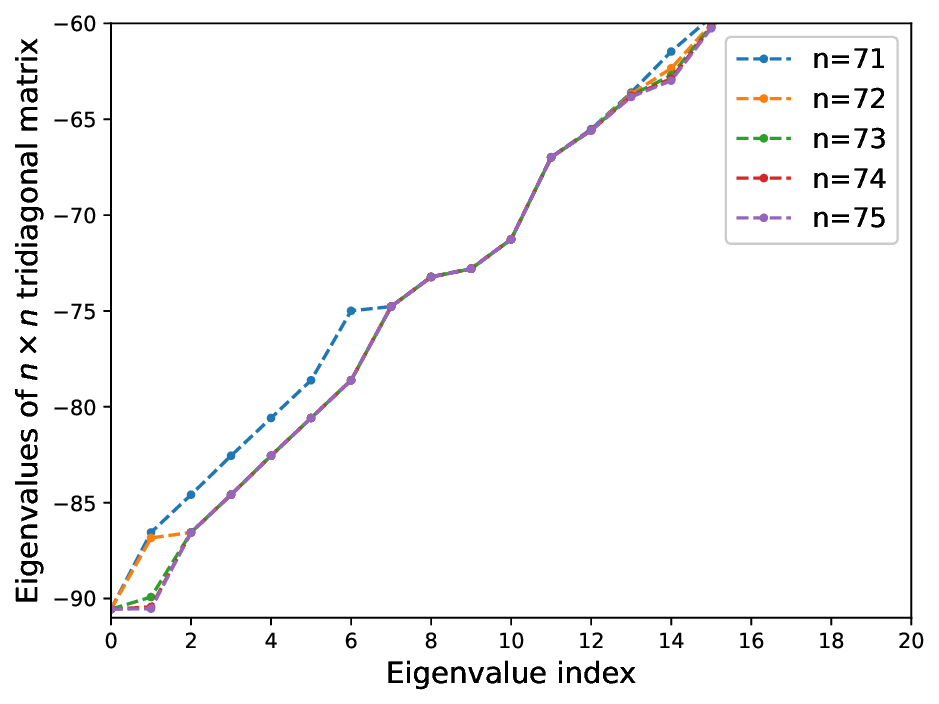}
\caption{\label{fig: e(t)}Extreme eigenvalues of truncated (at $n$) tri-diagonal matrices.}
\end{figure}

The Lanczos algorithm is known to be numerically unstable, and the nature of errors and alternative algorithms such as PRO and FO have been extensively studied in the literature \cite{Paige1971_thesis_errorLanczos, Parlett1979_Lanczos_SO, Simon1984_Lanczos_PRO}. But degenerate matrices introduce a new source of error that these algorithms fail to overcome. In the case of a degenerate matrix, the Krylov vectors could accumulate numerical error and leak out of the actual/analytical Krylov space into the rest of the much larger computational space, which is not improvable by alternate algorithms like PRO or FO \cite{k94p-vls8_error_Escaping_Krylovspace}. Ideally, each eigenspace (of dimension $p$) of $\mathcal{L}$ contributes one Krylov vector, leaving behind $p-1$ orthogonal directions that the next Krylov vectors could accumulate a component along. The Lanczos algorithm could then revisit the same eigenspace, as reflected in the eigenvalues of the truncated $n\times n$ tridiagonal matrices, in the form of multiple eigenvalues (see Fig. \ref{fig: e(t)}).

The eigenvalues of the truncated $n\times n $ tri-diagonal matrix converge to those of the full matrix, starting from the extreme values. As we continue, if the vectors start leaking out of the Krylov subspace, the largest eigenvalue will become degenerate, and we stop the numerics a few iterations earlier.

\section{Symmetries of the Hamiltonian}\label{appndx: symmetries of H}

In this appendix, we briefly review the continuous and discrete symmetries of the Uncoloured Tensor Model ($n=3,d=3$), which are relevant to the main text.

\subsection{Continuous Symmetries}
The Lagrangian of the Uncoloured Tensor Model is invariant under independent $O(3)$ rotations on each tensor index,
\begin{equation}
    \psi^{ijk} \rightarrow M_1^{ii'} M_2^{jj'} M_3^{kk'} \psi^{i'j'k'} \, , \quad \,M_1,M_2,M_3 \in O(3)
\end{equation}
The corresponding Noether charges are
\begin{align}
    Q_{1}^{ab}&=i\,\psi^{ajk}\psi^{bjk},\\
    Q_{2}^{ab}&=i\,\psi^{iak}\psi^{ibk},\\
    Q_{3}^{ab}&=i\,\psi^{ija}\psi^{ijb},
\end{align}
where  $a\neq b\in\{1,2,3\}$.

\subsection{Discrete Symmetries}
The Hamiltonian also has numerous discrete symmetries. The eigenvalue spectrum of the  Hamiltonian shows a mirror symmetry about a mid-value and has 16-fold degeneracy. These have been explained in terms of several discrete symmetries of the Hamiltonian in \cite{Krishnan2017_uncolored}. The symmetry operators identified in \cite{Krishnan2017_uncolored} are the following:
\begin{itemize}
\item Fermion parity operator,
\begin{equation}
    P=(-i)^{14} \gamma_1 \gamma_2\,...\gamma_{28}.
\end{equation}
\item The $Z$ operator,
\begin{equation}
   Z=(-i)\gamma_1\gamma_2\,...\gamma_{27} .
\end{equation} 
\item Time-reversal operator,
\begin{equation}
    T=-\gamma_2\gamma_4\,...\gamma_{28}\,\mathcal{K}.
\end{equation} 
\item The $U$ operator,
\begin{equation}
    U=\gamma_2\gamma_4\gamma_6\,...\gamma_{26}.
\end{equation} 
\item Fermion number operator,
\begin{equation}
    Q=\sum_{i=1}^{14} c_i^\dagger c_i,
\end{equation}
where
\begin{equation}
    c_a= \frac{1}{2}(\gamma_{2a-1}+ i\gamma_{2a}),\quad
    c_a^\dagger = \frac{1}{2}(\gamma_{2a-1}- i\gamma_{2a}).
\end{equation}
While the fermion parity number is not conserved, $(-1)^Q$ is a symmetry of $H$. 
\item The following operators commute with $H$,
\begin{align}
    N_1&=\gamma_1\gamma_2\,...\gamma_9\\
    N_2&=\gamma_{10}\gamma_{11}\,...\gamma_{18}\\
    N_3&=\gamma_{19}\gamma_{20}\,...\gamma_{27}
\end{align}

These charges obey the following commutation relations:
\begin{equation}\label{eq:discrete_algebra}
\begin{aligned}
     &[P, T] = [P, Q] = 0, \\
     &[Z, U] = [Z, Q] = [Z, N_i] = 0, \\
     &[T, U] = [T, N_i] = 0, \\
     &[N_2, U] = [N_1 N_2, Q] = 0.
\end{aligned}
\end{equation}
and these anti-commutation relations:
\begin{equation}\label{eq:discrete_algebra}
\begin{aligned}
     & \{P, Z\} = \{P, U\} = \{P, N_i\} = 0 \\
     & \{Z, T\} = 0, \\
     & \{N_{1,3}, U\} = 0.
\end{aligned}
\end{equation}

\end{itemize}

\bibliographystyle{jhep}
\bibliography{references3.bib}

@article{PhysRevX.9.041017_universalopgrowthhypothesis_parkerEtAl,
  title = "{A Universal Operator Growth Hypothesis}",
  author = {Parker, Daniel E. and Cao, Xiangyu and Avdoshkin, Alexander and Scaffidi, Thomas and Altman, Ehud},
  journal = {Phys. Rev. X},
  volume = {9},
  issue = {4},
  pages = {041017},
  numpages = {29},
  year = {2019},
  month = {Oct},
  publisher = {American Physical Society},
  doi = {10.1103/PhysRevX.9.041017},
  url = {https://link.aps.org/doi/10.1103/PhysRevX.9.041017}
}

@article{Krishnan2017bala_colored,
  author    = {Chethan Krishnan and Sambuddha Sanyal and P. N. Bala Subramanian},
  title     = "{Quantum chaos and holographic tensor models}",
  journal   = {Journal of High Energy Physics},
  year      = {2017},
  volume    = {2017},
  number    = {3},
  pages     = {56},
  doi       = {10.1007/JHEP03(2017)056},
  url       = {https://doi.org/10.1007/JHEP03(2017)056},
  issn      = {1029-8479}
}

@article{Krishnan2017_uncolored,
  author    = {Chethan Krishnan and K. V. Pavan Kumar and Sambuddha Sanyal},
  title     = "{Random matrices and holographic tensor models}",
  journal   = {Journal of High Energy Physics},
  year      = {2017},
  volume    = {2017},
  number    = {6},
  pages     = {36},
  doi       = {10.1007/JHEP06(2017)036},
  url       = {https://doi.org/10.1007/JHEP06(2017)036},
  issn      = {1029-8479}
}

@Article{Jian2021,
author={Jian, Shao-Kai
and Swingle, Brian
and Xian, Zhuo-Yu},
title="{Complexity growth of operators in the SYK model and in JT gravity}",
journal={Journal of High Energy Physics},
year={2021},
month={Mar},
day={01},
volume={2021},
number={3},
pages={14},
issn={1029-8479},
doi={10.1007/JHEP03(2021)014},
url={https://doi.org/10.1007/JHEP03(2021)014}
}

@Article{Maldacena2015waaAboundonchaos,
author={Maldacena, Juan
and Shenker, Stephen H.
and Stanford, Douglas},
title="{A bound on chaos}",
journal={Journal of High Energy Physics},
year={2016},
month={Aug},
day={17},
volume={2016},
number={8},
pages={106},
issn={1029-8479},
doi={10.1007/JHEP08(2016)106},
url={https://doi.org/10.1007/JHEP08(2016)106}
}

@article{PhysRevLett.70.3339_SYKintro,
  title = "{Gapless spin-fluid ground state in a random quantum Heisenberg magnet}",
  author = {Sachdev, Subir and Ye, Jinwu},
  journal = {Phys. Rev. Lett.},
  volume = {70},
  issue = {21},
  pages = {3339--3342},
  numpages = {0},
  year = {1993},
  month = {May},
  publisher = {American Physical Society},
  doi = {10.1103/PhysRevLett.70.3339},
  url = {https://link.aps.org/doi/10.1103/PhysRevLett.70.3339}
}

@ARTICLE{2016arXiv161009758W_edwardwitten,
       author = {{Witten}, Edward},
        title = "{An SYK-Like Model Without Disorder}",
      journal = {arXiv e-prints},
         year = 2016,
        month = oct,
          eid = {arXiv:1610.09758},
        pages = {arXiv:1610.09758},
          doi = {10.48550/arXiv.1610.09758},
archivePrefix = {arXiv},
       eprint = {1610.09758},
 primaryClass = {hep-th},
       adsurl = {https://ui.adsabs.harvard.edu/abs/2016arXiv161009758W}
}

@misc{Kitaev_lectureonline,
  author       = {Alexei Kitaev},
  title        = "{A simple model of quantum holography}",
  howpublished = {\url{https://online.kitp.ucsb.edu/online/entangled15/kitaev/}},
  note         = {KITP strings seminar and Entanglement 2015 program},
  year         = {2015},
  month        = {May},
  day          = {27},
}

@article{PhysRevD.95.046004_uncolored_model_into,
  title = "{Uncolored random tensors, melon diagrams, and the Sachdev-Ye-Kitaev models}",
  author = {Klebanov, Igor R. and Tarnopolsky, Grigory},
  journal = {Phys. Rev. D},
  volume = {95},
  issue = {4},
  pages = {046004},
  numpages = {13},
  year = {2017},
  month = {Feb},
  publisher = {American Physical Society},
  doi = {10.1103/PhysRevD.95.046004},
  url = {https://link.aps.org/doi/10.1103/PhysRevD.95.046004}
}

@article{Krishnan2018_conrastingSYKlike,
  author       = {Chethan Krishnan and K. V. Pavan Kumar and Dario Rosa},
  title        = "{Contrasting SYK-like models}",
  journal      = {Journal of High Energy Physics},
  year         = {2018},
  volume       = {2018},
  number       = {1},
  pages        = {64},
  doi          = {10.1007/JHEP01(2018)064},
  url          = {https://doi.org/10.1007/JHEP01(2018)064},
  issn         = {1029-8479}
}

@article{Maldacena_2016_commentsonthesykmodel_ig:OTOC,
  title = "{Remarks on the Sachdev-Ye-Kitaev model}",
  author = {Maldacena, Juan and Stanford, Douglas},
  journal = {Phys. Rev. D},
  volume = {94},
  issue = {10},
  pages = {106002},
  numpages = {43},
  year = {2016},
  month = {Nov},
  publisher = {American Physical Society},
  doi = {10.1103/PhysRevD.94.106002},
  url = {https://link.aps.org/doi/10.1103/PhysRevD.94.106002}
}

@article{PhysRevD.94.126010_spectralRandomMatrixSYK_levelstat,
  title = "{Spectral and thermodynamic properties of the Sachdev-Ye-Kitaev model}",
  author = {Garc\'{\i}a-Garc\'{\i}a, Antonio M. and Verbaarschot, Jacobus J. M.},
  journal = {Phys. Rev. D},
  volume = {94},
  issue = {12},
  pages = {126010},
  numpages = {13},
  year = {2016},
  month = {Dec},
  publisher = {American Physical Society},
  doi = {10.1103/PhysRevD.94.126010},
  url = {https://link.aps.org/doi/10.1103/PhysRevD.94.126010}
}

@article{Cotler2017_Blackholesandrandommatrices_SYK_SSF,
  author = {Cotler, Jordan S. and Gur-Ari, Guy and Hanada, Masanori and Polchinski, Joseph and Saad, Phil and Shenker, Stephen H. and Stanford, Douglas and Streicher, Alexandre and Tezuka, Masaki},
  title = "{Black holes and random matrices}",
  journal = {Journal of High Energy Physics},
  volume = {2017},
  number = {5},
  pages = {118},
  year = {2017},
  month = {05},
  day = {22},
  doi = {10.1007/JHEP05(2017)118},
  issn = {1029-8479}
}

@article{Rabinovici2021_Kcomp_thesis_I,
  author       = {Rabinovici, E. and Sánchez-Garrido, A. and Shir, R. and Sonner, J.},
  title        = "{Operator complexity: a journey to the edge of Krylov space}",
  journal      = {Journal of High Energy Physics},
  year         = {2021},
  volume       = {2021},
  number       = {6},
  pages        = {62},
  doi          = {10.1007/JHEP06(2021)062},
  url          = {https://doi.org/10.1007/JHEP06(2021)062}
}

@article{9v9v-54zv_blockdiagKrylov,
  title = "{Growth of block-diagonal operators and symmetry-resolved Krylov complexity}",
  author = {Caputa, Pawel and Di Giulio, Giuseppe and Loc, Tran Quang},
  journal = {Phys. Rev. Res.},
  volume = {7},
  issue = {4},
  pages = {043055},
  numpages = {15},
  year = {2025},
  month = {Oct},
  publisher = {American Physical Society},
  doi = {10.1103/9v9v-54zv},
  url = {https://link.aps.org/doi/10.1103/9v9v-54zv}
}

@phdthesis{Paige1971_thesis_errorLanczos,
  author  = {Christopher Conway Paige},
  title   = "{The Computation of Eigenvalues and Eigenvectors of Very Large Sparse Matrices}",
  school  = {University of London},
  address = {London, UK},
  year    = {1971},
  note    = {Thesis submitted for the degree of Doctor of Philosophy, London University Institute of Computer Science}
}

@article{Parlett1979_Lanczos_SO,
  title = "{The Lanczos Algorithm with Selective Orthogonalization}",
  author = {Parlett, B. N. and Scott, D. S.},
  journal = {Math. Comp.},
  volume = {33},
  number = {145},
  pages = {217--238},
  year = {1979},
  doi = {10.2307/2006037},
  url = {https://doi.org/10.2307/2006037}
}

@article{Simon1984_Lanczos_PRO,
  author  = {Horst D. Simon},
  title   = "{The Lanczos Algorithm with Partial Reorthogonalization}",
  journal = {Mathematics of Computation},
  volume  = {42},
  number  = {165},
  pages   = {115--142},
  year    = {1984},
  doi     = {10.2307/2007563},
  url     = {https://doi.org/10.2307/2007563}
}

@article{k94p-vls8_error_Escaping_Krylovspace,
  title = "{Escaping the Krylov space during the finite-precision Lanczos algorithm}",
  author = {Eckseler, Jannis and Pieper, Max and Schnack, J\"urgen},
  journal = {Phys. Rev. E},
  volume = {112},
  issue = {2},
  pages = {025306},
  numpages = {7},
  year = {2025},
  month = {Aug},
  publisher = {American Physical Society},
  doi = {10.1103/k94p-vls8},
  url = {https://link.aps.org/doi/10.1103/k94p-vls8}
}

@book{viswanath_2025_avfvz-v0s16,
  author       = {Viswanath, V.S. and
                  Mueller, G.},
  title        = "{The recursion method. Application to many-body dynamics}",
  publisher    = {Springer.},
  year         = 2025,
  volume       = 23,
  month        = jan,
}

@article{Hawking:1982dh,
    author = "Hawking, S. W. and Page, Don N.",
    title = "{Thermodynamics of Black Holes in anti-De Sitter Space}",
    reportNumber = "PRINT-83-0019 (CAMBRIDGE)",
    doi = "10.1007/BF01208266",
    journal = "Commun. Math. Phys.",
    volume = "87",
    pages = "577",
    year = "1983"
}

@article{Bekenstein:1973ur,
    author = "Bekenstein, Jacob D.",
    title = "{Black holes and entropy}",
    doi = "10.1103/PhysRevD.7.2333",
    journal = "Phys. Rev. D",
    volume = "7",
    pages = "2333--2346",
    year = "1973"
}

@article{tHooft:1993dmi,
    author = "'t Hooft, Gerard",
    title = "{Dimensional reduction in quantum gravity}",
    eprint = "gr-qc/9310026",
    archivePrefix = "arXiv",
    reportNumber = "THU-93-26",
    journal = "Conf. Proc. C",
    volume = "930308",
    pages = "284--296",
    year = "1993"
}

@article{Susskind:1994vu,
    author = "Susskind, Leonard",
    title = "{The World as a hologram}",
    eprint = "hep-th/9409089",
    archivePrefix = "arXiv",
    reportNumber = "SU-ITP-94-33",
    doi = "10.1063/1.531249",
    journal = "J. Math. Phys.",
    volume = "36",
    pages = "6377--6396",
    year = "1995"
}

@article{Hawking:1975vcx,
    author = "Hawking, S. W.",
    editor = "Gibbons, G. W. and Hawking, S. W.",
    title = "{Particle Creation by Black Holes}",
    doi = "10.1007/BF02345020",
    journal = "Commun. Math. Phys.",
    volume = "43",
    pages = "199--220",
    year = "1975",
    note = "[Erratum: Commun.Math.Phys. 46, 206 (1976)]"
}

@article{Maldacena:1997re,
    author = "Maldacena, Juan Martin",
    title = "{The Large $N$ limit of superconformal field theories and supergravity}",
    eprint = "hep-th/9711200",
    archivePrefix = "arXiv",
    reportNumber = "HUTP-97-A097, HUTP-98-A097",
    doi = "10.4310/ATMP.1998.v2.n2.a1",
    journal = "Adv. Theor. Math. Phys.",
    volume = "2",
    pages = "231--252",
    year = "1998"
}

@article{Witten:1998qj,
    author = "Witten, Edward",
    title = "{Anti de Sitter space and holography}",
    eprint = "hep-th/9802150",
    archivePrefix = "arXiv",
    reportNumber = "IASSNS-HEP-98-15",
    doi = "10.4310/ATMP.1998.v2.n2.a2",
    journal = "Adv. Theor. Math. Phys.",
    volume = "2",
    pages = "253--291",
    year = "1998"
}

@article{Herzog_2007,
   title="{Quantum critical transport, duality, and M theory}",
   volume={75},
   ISSN={1550-2368},
   url={http://dx.doi.org/10.1103/PhysRevD.75.085020},
   DOI={10.1103/physrevd.75.085020},
   number={8},
   journal={Physical Review D},
   publisher={American Physical Society (APS)},
   author={Herzog, Christopher P. and Kovtun, Pavel and Sachdev, Subir and Son, Dam Thanh},
   year={2007},
   month=apr }

@article{Gurau_2017,
   title="{The complete 1/N expansion of a SYK–like tensor model}",
   volume={916},
   ISSN={0550-3213},
   url={http://dx.doi.org/10.1016/j.nuclphysb.2017.01.015},
   DOI={10.1016/j.nuclphysb.2017.01.015},
   journal={Nuclear Physics B},
   publisher={Elsevier BV},
   author={Gurau, Razvan},
   year={2017},
   month=mar, pages={386–401} }

@article{Caputa:2021sib,
    author = "Caputa, Pawel and Magan, Javier M. and Patramanis, Dimitrios",
    title = "{Geometry of Krylov complexity}",
    eprint = "2109.03824",
    archivePrefix = "arXiv",
    primaryClass = "hep-th",
    doi = "10.1103/PhysRevResearch.4.013041",
    journal = "Phys. Rev. Res.",
    volume = "4",
    number = "1",
    pages = "013041",
    year = "2022"
}

@article{Shenker:2013pqa,
    author = "Shenker, Stephen H. and Stanford, Douglas",
    title = "{Black holes and the butterfly effect}",
    eprint = "1306.0622",
    archivePrefix = "arXiv",
    primaryClass = "hep-th",
    reportNumber = "SU-ITP-13-08",
    doi = "10.1007/JHEP03(2014)067",
    journal = "JHEP",
    volume = "03",
    pages = "067",
    year = "2014"
}

@article{Rabinovici_2022,
   title="{Krylov complexity from integrability to chaos}",
   volume={2022},
   ISSN={1029-8479},
   url={http://dx.doi.org/10.1007/JHEP07(2022)151},
   DOI={10.1007/jhep07(2022)151},
   number={7},
   journal={Journal of High Energy Physics},
   publisher={Springer Science and Business Media LLC},
   author={Rabinovici, E. and Sánchez-Garrido, A. and Shir, R. and Sonner, J.},
   year={2022},
   month=jul }

@article{Bhattacharyya_2023,
   title= "{Operator growth and Krylov complexity in Bose-Hubbard model}",
   volume={2023},
   ISSN={1029-8479},
   url={http://dx.doi.org/10.1007/JHEP12(2023)112},
   DOI={10.1007/jhep12(2023)112},
   number={12},
   journal={Journal of High Energy Physics},
   publisher={Springer Science and Business Media LLC},
   author={Bhattacharyya, Arpan and Ghosh, Debodirna and Nandi, Poulami},
   year={2023},
   month=dec }

@article{Noh_2021,
   title="{Operator growth in the transverse-field Ising spin chain with integrability-breaking longitudinal field}",
   volume={104},
   ISSN={2470-0053},
   url={http://dx.doi.org/10.1103/PhysRevE.104.034112},
   DOI={10.1103/physreve.104.034112},
   number={3},
   journal={Physical Review E},
   publisher={American Physical Society (APS)},
   author={Noh, Jae Dong},
   year={2021},
   month=sep }

@article{Yates_2020,
   title="{Dynamics of almost strong edge modes in spin chains away from integrability}",
   volume={102},
   ISSN={2469-9969},
   url={http://dx.doi.org/10.1103/PhysRevB.102.195419},
   DOI={10.1103/physrevb.102.195419},
   number={19},
   journal={Physical Review B},
   publisher={American Physical Society (APS)},
   author={Yates, Daniel J. and Abanov, Alexander G. and Mitra, Aditi},
   year={2020},
   month=nov }

@article{Bhattacharya_2024,
   title="{Krylov complexity for nonlocal spin chains}",
   volume={109},
   ISSN={2470-0029},
   url={http://dx.doi.org/10.1103/PhysRevD.109.066010},
   DOI={10.1103/physrevd.109.066010},
   number={6},
   journal={Physical Review D},
   publisher={American Physical Society (APS)},
   author={Bhattacharya, Aranya and Nath, Pingal Pratyush and Sahu, Himanshu},
   year={2024},
   month=mar }

@article{neumann1929beweis,
  title="{Beweis des Ergodensatzes und des H-Theorems in der neuen Mechanik}",
  author={Neumann, J v},
  journal={Zeitschrift f{\"u}r Physik},
  volume={57},
  number={1},
  pages={30--70},
  year={1929},
  publisher={Springer}
}

@ARTICLE{2006NatPh...2..754P,
       author = {{Popescu}, Sandu and {Short}, Anthony J. and {Winter}, Andreas},
        title = "{Entanglement and the foundations of statistical mechanics}",
      journal = {Nature Physics},
         year = 2006,
        month = nov,
       volume = {2},
       number = {11},
        pages = {754-758},
          doi = {10.1038/nphys444},
archivePrefix = {arXiv},
       eprint = {quant-ph/0511225},
 primaryClass = {quant-ph},
       adsurl = {https://ui.adsabs.harvard.edu/abs/2006NatPh...2..754P}
}

@article{Srednicki_1994,
   title="{Chaos and quantum thermalization}",
   volume={50},
   ISSN={1095-3787},
   url={http://dx.doi.org/10.1103/PhysRevE.50.888},
   DOI={10.1103/physreve.50.888},
   number={2},
   journal={Physical Review E},
   publisher={American Physical Society (APS)},
   author={Srednicki, Mark},
   year={1994},
   month=aug, pages={888–901} }

@article{Zeh1970,
  author = {Zeh, H. D.},
  title ="{On the interpretation of measurement in quantum theory}",
  journal = {Foundations of Physics},
  year = {1970},
  volume = {1},
  number = {1},
  pages = {69--76},
  doi = {10.1007/BF00708656},
  url = {https://link.springer.com/article/10.1007/BF00708656}
}

@book{Schlosshauer2007decoherence,
  author    = {Schlosshauer, Maximilian A.},
  title     = "{Decoherence and the Quantum-to-Classical Transition}",
  series    = {The frontiers collection},
  publisher = {Springer},
  address   = {Berlin, Germany},
  year      = {2007},
  isbn      = {978-3-540-35773-5},
  doi       = {10.1007/978-3-540-35775-9}
}

@article{Geloun_2017,
   title="{Tensor models, Kronecker coefficients and permutation centralizer algebras}",
   volume={2017},
   ISSN={1029-8479},
   url={http://dx.doi.org/10.1007/JHEP11(2017)092},
   DOI={10.1007/jhep11(2017)092},
   number={11},
   journal={Journal of High Energy Physics},
   publisher={Springer Science and Business Media LLC},
   author={Geloun, Joseph Ben and Ramgoolam, Sanjaye},
   year={2017},
   month=nov }

@article{Itoyama_2017,
   title="{Rainbow tensor model with enhanced symmetry and extreme melonic dominance}",
   volume={771},
   ISSN={0370-2693},
   url={http://dx.doi.org/10.1016/j.physletb.2017.05.043},
   DOI={10.1016/j.physletb.2017.05.043},
   journal={Physics Letters B},
   publisher={Elsevier BV},
   author={Itoyama, H. and Mironov, A. and Morozov, A.},
   year={2017},
   month=aug, pages={180–188} }

@article{Chaudhuri_2018,
   title="{Abelian tensor models on the lattice}",
   volume={97},
   ISSN={2470-0029},
   url={http://dx.doi.org/10.1103/PhysRevD.97.086007},
   DOI={10.1103/physrevd.97.086007},
   number={8},
   journal={Physical Review D},
   publisher={American Physical Society (APS)},
   author={Chaudhuri, Soumyadeep and Giraldo-Rivera, Victor I. and Joseph, Anosh and Loganayagam, R. and Yoon, Junggi},
   year={2018},
   month=apr }

@article{Choudhury_2018,
   title="{Notes on melonic $O(N)^{q-1}$ tensor models}",
   volume={2018},
   ISSN={1029-8479},
   url={http://dx.doi.org/10.1007/JHEP06(2018)094},
   DOI={10.1007/jhep06(2018)094},
   number={6},
   journal={Journal of High Energy Physics},
   publisher={Springer Science and Business Media LLC},
   author={Choudhury, Sayantan and Dey, Anshuman and Halder, Indranil and Janagal, Lavneet and Minwalla, Shiraz and Poojary, Rohan R.},
   year={2018},
   month=jun }

@article{Carrozza_2018,
   title="{Large N limit of irreducible tensor models: O(N) rank-3 tensors with mixed permutation symmetry}",
   volume={2018},
   ISSN={1029-8479},
   url={http://dx.doi.org/10.1007/JHEP06(2018)039},
   DOI={10.1007/jhep06(2018)039},
   number={6},
   journal={Journal of High Energy Physics},
   publisher={Springer Science and Business Media LLC},
   author={Carrozza, Sylvain},
   year={2018},
   month=jun }

@article{PhysRevLett.120.201603,
  title = "{Exact Solution of a Strongly Coupled Gauge Theory in $0+1$ Dimensions}",
  author = {Krishnan, Chethan and Kumar, K. V. Pavan},
  journal = {Phys. Rev. Lett.},
  volume = {120},
  issue = {20},
  pages = {201603},
  numpages = {5},
  year = {2018},
  month = {May},
  publisher = {American Physical Society},
  doi = {10.1103/PhysRevLett.120.201603},
  url = {https://link.aps.org/doi/10.1103/PhysRevLett.120.201603}
}

@article{Krishnan_2017,
   title="{Towards a finite-N hologram}",
   volume={2017},
   ISSN={1029-8479},
   url={http://dx.doi.org/10.1007/JHEP10(2017)099},
   DOI={10.1007/jhep10(2017)099},
   number={10},
   journal={Journal of High Energy Physics},
   publisher={Springer Science and Business Media LLC},
   author={Krishnan, Chethan and Kumar, K. V. Pavan},
   year={2017},
   month=oct }

@article{Klebanov_2018,
   title="{Spectra of eigenstates in fermionic tensor quantum mechanics}",
   volume={97},
   ISSN={2470-0029},
   url={http://dx.doi.org/10.1103/PhysRevD.97.106023},
   DOI={10.1103/physrevd.97.106023},
   number={10},
   journal={Physical Review D},
   publisher={American Physical Society (APS)},
   author={Klebanov, Igor R. and Milekhin, Alexey and Popov, Fedor and Tarnopolsky, Grigory},
   year={2018},
   month=may }

@article{Hashimoto_2023,
   title="{Krylov complexity and chaos in quantum mechanics}",
   volume={2023},
   ISSN={1029-8479},
   url={http://dx.doi.org/10.1007/JHEP11(2023)040},
   DOI={10.1007/jhep11(2023)040},
   number={11},
   journal={Journal of High Energy Physics},
   publisher={Springer Science and Business Media LLC},
   author={Hashimoto, Koji and Murata, Keiju and Tanahashi, Norihiro and Watanabe, Ryota},
   year={2023},
   month=nov }

@article{Avdoshkin_2024,
   title="{Krylov complexity in quantum field theory, and beyond}",
   volume={2024},
   ISSN={1029-8479},
   url={http://dx.doi.org/10.1007/JHEP06(2024)066},
   DOI={10.1007/jhep06(2024)066},
   number={6},
   journal={Journal of High Energy Physics},
   publisher={Springer Science and Business Media LLC},
   author={Avdoshkin, Alexander and Dymarsky, Anatoly and Smolkin, Michael},
   year={2024},
   month=jun }

@article{Baggioli_2025,
   title="{Krylov complexity as an order parameter for quantum chaotic-integrable transitions}",
   volume={7},
   ISSN={2643-1564},
   url={http://dx.doi.org/10.1103/PhysRevResearch.7.023028},
   DOI={10.1103/physrevresearch.7.023028},
   number={2},
   journal={Physical Review Research},
   publisher={American Physical Society (APS)},
   author={Baggioli, Matteo and Huh, Kyoung-Bum and Jeong, Hyun-Sik and Kim, Keun-Young and Pedraza, Juan F.},
   year={2025},
   month=apr }

@misc{rabinovici2025krylovcomplexity,
      title="{Krylov Complexity}", 
      author={Eliezer Rabinovici and Adrián Sánchez-Garrido and Ruth Shir and Julian Sonner},
      year={2025},
      eprint={2507.06286},
      archivePrefix={arXiv},
      primaryClass={hep-th},
      url={https://arxiv.org/abs/2507.06286}, 
}

@article{pavan,
   title="{(Anti-)symmetrizing wave functions}",
   volume={60},
   ISSN={1089-7658},
   url={http://dx.doi.org/10.1063/1.5038076},
   DOI={10.1063/1.5038076},
   number={2},
   journal={Journal of Mathematical Physics},
   publisher={AIP Publishing},
   author={Krishnan, Chethan and Kumar, K. V. Pavan and Subramanian, P. N. Bala},
   year={2019},
   month=feb }

@misc{maldacena2016conformalsymmetrybreakingdimensional,
      title="{Conformal symmetry and its breaking in two dimensional Nearly Anti-de-Sitter space}", 
      author={Juan Maldacena and Douglas Stanford and Zhenbin Yang},
      year={2016},
      eprint={1606.01857},
      archivePrefix={arXiv},
      primaryClass={hep-th},
      url={https://arxiv.org/abs/1606.01857}, 
}

@article{Carrozza:2015adg,
    author = "Carrozza, Sylvain and Tanasa, Adrian",
    title = "{$O(N)$ Random Tensor Models}",
    eprint = "1512.06718",
    archivePrefix = "arXiv",
    primaryClass = "math-ph",
    doi = "10.1007/s11005-016-0879-x",
    journal = "Lett. Math. Phys.",
    volume = "106",
    number = "11",
    pages = "1531--1559",
    year = "2016"
}

@article{Caputa_2024,
   title="{Krylov complexity of density matrix operators}",
   volume={2024},
   ISSN={1029-8479},
   url={http://dx.doi.org/10.1007/JHEP05(2024)337},
   DOI={10.1007/jhep05(2024)337},
   number={5},
   journal={Journal of High Energy Physics},
   publisher={Springer Science and Business Media LLC},
   author={Caputa, Pawel and Jeong, Hyun-Sik and Liu, Sinong and Pedraza, Juan F. and Qu, Le-Chen},
   year={2024},
   month=may }

@article{PhysRevD.109.046017,
  title = "{Spectral and Krylov complexity in billiard systems}",
  author = {Camargo, Hugo A. and Jahnke, Viktor and Jeong, Hyun-Sik and Kim, Keun-Young and Nishida, Mitsuhiro},
  journal = {Phys. Rev. D},
  volume = {109},
  issue = {4},
  pages = {046017},
  numpages = {20},
  year = {2024},
  month = {Feb},
  publisher = {American Physical Society},
  doi = {10.1103/PhysRevD.109.046017},
  url = {https://link.aps.org/doi/10.1103/PhysRevD.109.046017}
}

@misc{chryssanthacopoulos2026krylovcomplexitysupersymmetricsyk,
      title="{Krylov Complexity of Supersymmetric SYK Models}", 
      author={James Chryssanthacopoulos and David Vegh},
      year={2026},
      eprint={2511.20769},
      archivePrefix={arXiv},
      primaryClass={hep-th},
      url={https://arxiv.org/abs/2511.20769}, 
}

@article{Nandy_2025,
   title="{Quantum dynamics in Krylov space: Methods and applications}",
   volume={1125–1128},
   ISSN={0370-1573},
   url={http://dx.doi.org/10.1016/j.physrep.2025.05.001},
   DOI={10.1016/j.physrep.2025.05.001},
   journal={Physics Reports},
   publisher={Elsevier BV},
   author={Nandy, Pratik and Matsoukas-Roubeas, Apollonas S. and Martínez-Azcona, Pablo and Dymarsky, Anatoly and del Campo, Adolfo},
   year={2025},
   month=jun, pages={1–82} }

@article{Bhattacharjee_2023,
   title="{Krylov complexity in large q and double-scaled SYK model}",
   volume={2023},
   ISSN={1029-8479},
   url={http://dx.doi.org/10.1007/JHEP08(2023)099},
   DOI={10.1007/jhep08(2023)099},
   number={8},
   journal={Journal of High Energy Physics},
   publisher={Springer Science and Business Media LLC},
   author={Bhattacharjee, Budhaditya and Nandy, Pratik and Pathak, Tanay},
   year={2023},
   month=aug }

@article{Caputa_2026,
   title="{Symmetry-resolved spread complexity}",
   volume={2026},
   ISSN={1029-8479},
   url={http://dx.doi.org/10.1007/JHEP02(2026)189},
   DOI={10.1007/jhep02(2026)189},
   number={2},
   journal={Journal of High Energy Physics},
   publisher={Springer Science and Business Media LLC},
   author={Caputa, Pawel and Di Giulio, Giuseppe and Loc, Tran Quang},
   year={2026},
   month=feb }
{}

\end{document}